\documentclass[aps,prb,twocolumn,groupedaddress,english]{revtex4-1}

\usepackage[T1]{fontenc}
\usepackage{babel}
\usepackage{amsmath}
\usepackage{amssymb}
\usepackage{wasysym}
\usepackage{graphicx}
\usepackage{xcolor}
\usepackage{graphicx}
\usepackage{braket}

\usepackage[linktocpage=true,
  colorlinks=true, 
  pdfborder={0 0 0},
  linkcolor=blue,
  citecolor=red,
  filecolor=yellow,
  urlcolor=blue,
  bookmarks,
  pdfauthor={},
]{hyperref}

\newcommand{\Graz}{Institute of Theoretical and Computational Physics, Graz University of Technology, NAWI Graz, 8010 Graz, Austria}
\newcommand{\sh}{SH$_3$}
\newcommand{\ph}{PH$_3$}

\newcommand{\tc}{$T_\text{c}$}

\newcommand{\omlog}{$\omega_{\textmd log}$}

\def\bb{\boldsymbol}

\begin{document}

\title{Prediction of High-$\bb{T}_\text{\!c}$ conventional Superconductivity in the Ternary Lithium Borohydride System}

\author{Christian Kokail} \email{ckokail@alumni.tugraz.at}
\author{Lilia Boeri}         
\author{Wolfgang von der Linden}   
\affiliation{\Graz}

\date{\today}

\begin{abstract}
We investigate the superconducting ternary lithium borohydride phase diagram at pressures of 0 and 200$\,$GPa using methods for evolutionary crystal structure prediction and linear-response calculations for the electron-phonon coupling. Our calculations show that the ground state phase at ambient pressure, LiBH$_4$, stays in the $Pnma$ space group and remains a wide band-gap insulator at all pressures investigated. Other phases along the 1:1:$x$ Li:B:H line are also insulating. However, a full search of the ternary phase diagram at 200$\,$GPa revealed a metallic Li$_2$BH$_6$ phase, which is thermodynamically stable down to 100$\,$GPa. This {\em superhydride} phase, crystallizing in a $Fm\bar{3}m$ space group, is characterized by six-fold hydrogen-coordinated boron atoms occupying the $fcc$ sites of the unit cell. Due to strong hydrogen-boron bonding this phase displays a critical temperature of $\sim$ 100$\,$K between 100 and 200$\,$GPa. Our investigations confirm that ternary compounds used in hydrogen-storage applications are a suitable choice for observing high-$T_\text{c}$ conventional superconductivity in diamond anvil cell experiments, and suggest a viable route
to optimize the critical temperature of high-pressure hydrides.      
\end{abstract}

\pacs{~}
\maketitle

The pioneering prediction of N.W. Ashcroft that hydrogen, the lightest among all elements, could become a high-temperature (high-\tc) superconductor at high pressures, 
can be seen as the foundation of high-pressure superconductivity research~\cite{PhysRevLett.21.1748}. 
Hydrogen has long been subject to comprehensive theoretical and experimental 
investigations,~\cite{Wigner_JCP1935,PhysRevLett.21.1748,PhysRevLett.100.257001,PhysRevB.84.144515,PhysRevB.93.174308,Szniak20092053,eremets2016low,nat0028-0836} since it is expected to exhibit many fascinating properties, including a superconductor to superfluid phase transition.~\cite{babaev}
In January 2017, Dias and Silvera reported its metallization in a diamond anvil
 cell under a static pressure of 495$\,$GPa~\cite{Dias715}. The heated discussion that this claim has initiated among the experts testifies the relevance and the high actuality of this topic~\cite{goncharov2017comment,eremets_comment}. 
Besides pure hydrogen, it has been demonstrated that also metallic 
hydrides can become high-\tc~superconductors at much lower pressures than those required to metallize hydrogen. 
Impurities in the hydrogen matrix can influence the bonding properties, and cause a \textit{chemical precompression} on the H atoms~\cite{PhysRevLett.92.187002,yao2010silane,PhysRevLett.96.017006}. 
This idea has led to the prediction of novel high-pressure hydrides, with remarkable superconducting transition temperatures. 
The coronation of this predictions was the experimental discovery of
\sh, with critical temperatures as
high as 203 \,$K$ at 200$\,$GPa~\cite{DrozdovEremets_Nature2015,Troyan1303,Duan_SciRep2014}. In addition to 
being the current record-holder for superconductivity,
\sh\ is the first example of a completely unknown compound predicted from first principles.
A few months after \sh, high-\tc\; superconductivity was reported in a 
second superconducting hydride, \ph.\cite{Drozdov_PH3_arxiv2015,shamp_decomposition_2015,Fu_Ma_pnictogenH_2016,Flores_PH3_PRBR2016}
Besides these two known examples, other hydrides have been predicted to superconduct above liquid nitrogen temperature,
\cite{zurek2009little,yao2010silane,Wang_PNAS2012_CaH6,Li_SREP2015_YH6,Struzhkin_rev,Flores_H2O,Zurek_IodineH_JPCL2016} 
but, in general, the \tc's of binary hydrides are quite scattered and only
a few of them surpass the liquid nitrogen threshold.~\cite{kim_general_2010}

\sh\; has been the object of several {\em ab-initio} studies, which have established that
its record-high \tc~  is a consequence of high electron-phonon ($ep$) matrix elements enabled by the strong hydrogen-sulfur bonds, electronic van-Hove singularities at the Fermi level, and large vibrational frequencies of the hydrogen modes accompanied by large anharmonic effects~\cite{Duan_SciRep2014,SH_PRB-Mazin-2015,Heil-Boeri_PRB2015,Flores-Livas2016,PhysRevB.93.104526,doi:10.1063/1.4874158,PhysRevB.91.184511,PhysRevB.94.064507,PhysRevLett.114.157004}.
The first two aspects are intrinsically related to the $Im\bar{3}m$ high-pressure  structure of \sh,
which is a typical example of {\em forbidden chemistry},
i.e. a behavior, which typically occurs at high pressures,
 that defies the usual rules of chemistry.
In this structure sulfur forms three 90$^{\circ}$ covalent bonds with hydrogen,
which couple strongly to phonons. It has been shown that in binary hydrides 
the  formation of metallic covalent bonds, conducive to high-\tc superconductivity, 
 requires elements with electronegativities close to 
hydrogen.~\cite{SH_PRB-Mazin-2015,Heil-Boeri_PRB2015,Fu_Ma_pnictogenH_2016} 
It is coinceivable that also other atomic properties, such as valence, atomic radii, etc could have an influence on the high-pressure superconducting 
behavior of hydrides. 
Understanding how these properties could be tuned to increase the maximum
critical temperatures or decrease the pressure needed to induce high-\tc\; superconductivity represents a major step forward for the 
design of better superconductors.

Hydrogen storage research has shown that
{\em complex} (ternary or higher) hydrides often 
exhibit improved performances compared to simple hydrides, 
because by controlling the chemical composition it is possible to improve independently different properties, such as hydrogen density and activation barriers. 
~\cite{nature_review_hydrogen} 
The same flexibility could be exploited to 
improve the superconducting behavior at high pressures,
for example acting independently on the 
doping level 
and on the bonding characteristics to lower 
the metallization pressure or increase the maximum \tc. 

Given the large number of ternary hydrides, identifying suitable systems experimentally 
by trial and error is unfeasible. On the other hand, {\em ab-initio} methods for crystal structure prediction and 
thermodynamics, which 
led to the succesful prediction of \sh, ~\cite{Duan_SciRep2014}
can also be applied  to multinary phase diagrams. 
The phase space and, consequently, the computational cost are much larger
as compared to binary hydrides.
Therefore, it is not surprising that, although a few examples of
{\em ab-initio} studies of the phase diagrams of complex hydrides at
ambient pressure can be found in 
literature,~\cite{wolverton_advmat_2007}
to our knowledge there are no examples 
of similar studies for superconductivity at high pressures.

In this work, we explore {\em ab-initio}
the high-pressure superconducting phase diagram of a prototypical 
ternary system,
lithium-boron-hydrogen, combining methods for evolutionary crystal structure prediction with linear-response calculations of the electron-phonon ($ep$) coupling.~\cite{Baroni_RMP2001,QE_JPCM_2009} Our aim 
is to identify prospective high-\tc\; superconductors at high pressures.
 We show that an accurate sampling
of the whole phase diagram is needed to identify the high-\tc\; 
superconducting phases,
because these are found for compositions that are not {\em obvious} 
 in the sense that will be discussed below.

The lithium-boron-hydrogen system is very well characterized at ambient pressures,
 because the ground-state lithium borohydride (LiBH$_4$) is one of the best materials for hydrogen storage applications.
This compound  combines a weak (Li) and a strong (B) hydrogen former, and this permits to have at the same time a high hydrogen
density and a reasonable activation barrier for hydrogenation 
and dehydrogenation reactions;
due to the low masses of Li and B, not only the volumetric density, but also the gravimetric one are extremely 
high;~\cite{scidirzuet2003,vajo2005,zuettel2003,PhysRevB.69.245120,shinorimo2006,soulie2002,
gross2008,mauronacs2008} furthermore,
the existence of several possible hydriding and dehydriding reactions 
provides the possibility to control the H-content in experiments.\cite{orimo2005,kolmogorov_libh_2015}

Except for the boundary phases, the high-pressure phase diagram is unknown, 
but there are many reasons to believe that it could host high-\tc\; superconductors.
First of all, the very light masses of the three constituents 
imply that the average phonon frequencies of all compounds will be high,
which is intrinsically favorable to phonon-mediated superctivity.
In fact Li, B and the corresponding hydrides exhibit interesting superconducting properties under pressure,~\cite{tuorinemi2007,shimizuli2002,eremtsli2002,zurek2009little,PhysRevLett.106.015503,PhysRevLett.106.095502,naumov2015chemical,kokail_PhysRevB.94.060502,Eremets272,PhysRevB.65.172510,PhysRevLett.110.165504}  while the binary Li-B system hosts one of the first {\em ab-initio} predictions of novel superconductors.~\cite{kolmogorov_lib_PRB2006}
Furthermore, strong hydride formers, such as boron, form covalent or ionic bonds, which translate into large intrinsic $ep$ matrix elements, 
while weak hydrogen formers typically form metallic hydrides; combining the properties of the two elements, therefore
a ternary Li-B-H compound could behave as a "covalent metal",
similarly to \sh, already at much lower pressures. 
The many hydrogen-rich phases which are weakly metastable at ambient pressure
are ideal candidates for covalent metallic behavior (and superconductivity): 
in fact, they could be considered the ternary equivalent of
\sh, which 
is a hydrogen-rich phase obtained by the hydring reaction of SH$_2$ at 
high pressures. ~\cite{DrozdovEremets_Nature2015,Duan_SciRep2014} 

The aim of this work is to understand whether 
 any of the ternary Li-B-H compounds known at ambient pressure, or any new, 
still unknown composition, exhibit high-\tc\; superconductivity
in the Megabar range. We indeed identify a new high-\tc\; phase (Li$_2$BH$_6$), 
which, similarly to \sh, can be classified as a highly-symmetric covalent
metal. At 200 GPa, this compound exhibits a \tc\; of 80 K, i.e. lower
than \sh\; but, in contrast to other known hydrides, the
high-\tc\; behaviour persists down to 100 GPa.
We will argue that the possibility to lower the pressure for high-\tc\; compared
to binary hydrides is an intrinsic property of ternary (or higher) hydrides.

\begin{figure}[t]
\includegraphics[width=1.0\columnwidth,angle=0]{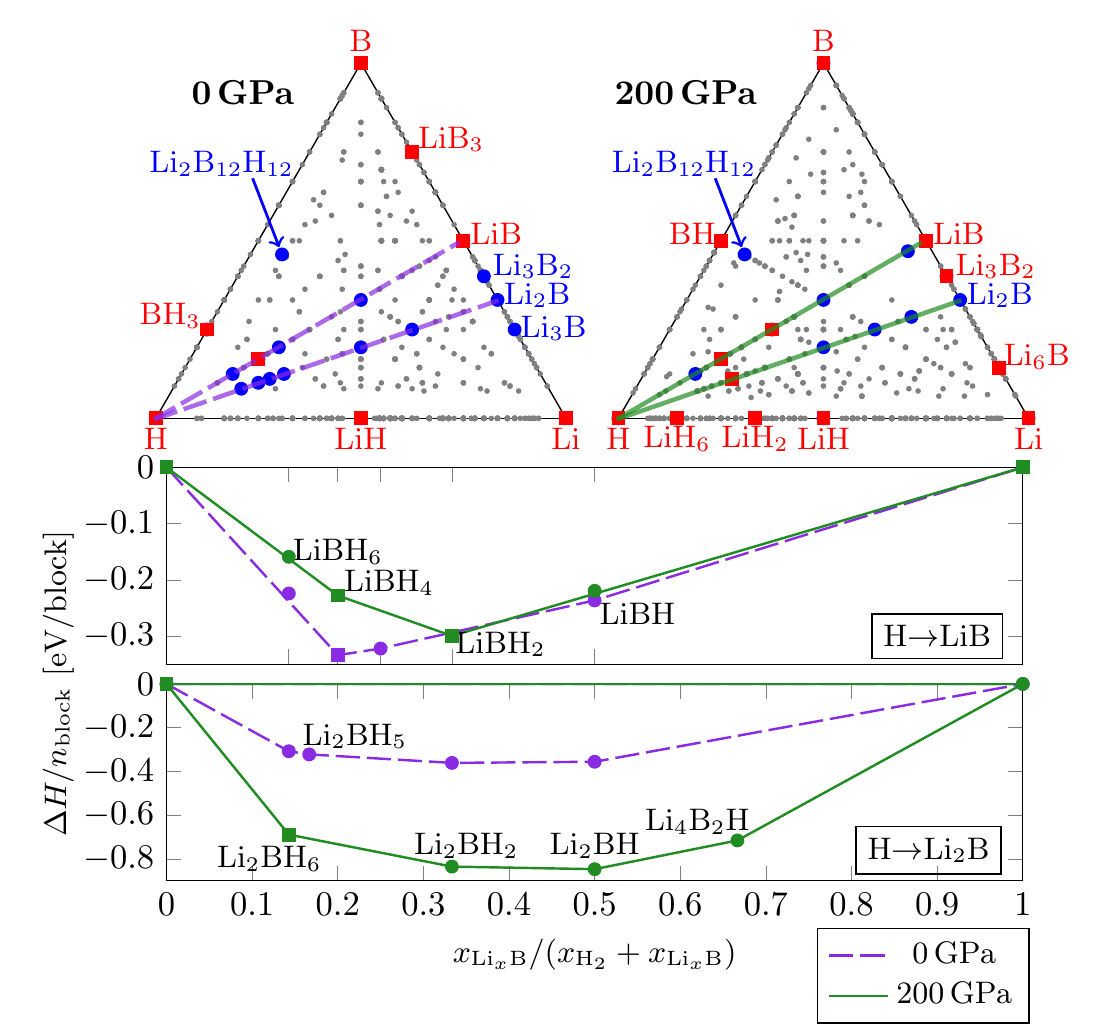}
\caption{{\em Top:} Generalized convex hull for
the Li-B-H system at zero (left) and 200 (right) GPa, obtained from
evolutionary crystal structure prediction. Points represent compositions
sampled in our preliminary run, lines indicate ranges of compositions
for which we computed more accurate binary convex hulls (see text).
These are shown in the {\em bottom panels}.
Circles and squares represent compositions that are thermodynamically metastable
or stable with respect to other phases on the ternary hull ({\em see text}).
}
\label{fig:TPD}
\end{figure}

Fig.~\ref{fig:TPD} shows the 
phase diagram of the Li-B-H system at ambient pressure ($P=0$) and at 
200 GPa ($P=200$),  calculated using the
evolutionary crystal structure prediction method, as implemented in the
the \textsc{Uspex} code.~\cite{uspex_note,oganov2006crystal,lyakhov2013new,oganov2011evolutionary}. 
Due to the high computational cost of ternary phase diagrams, 
we had to restrict our search to representative pressures:
200 GPa was chosen because this is the
pressure at which \sh\ exhibits its maximum \tc, and is well beyond
the metallization pressure for many binary hydrides. Ambient
pressure was mainly intended to check the accuracy of our
calculations against literature results. 

We first performed a full search of the two ternary phase diagrams, 
in which we sampled many possible compositions, represented
by symbols in the two upper panels. The aim of this preliminary 
scan is to identify the compositions in the ternary phase
diagram that could give rise to high-symmetry metallic
structures at high pressure. 
In order to ensure an optimal trade-off between accuracy and computational time, we
restricted the search to structures with all possible compositions, but with a minimum(maximum) 
number of atoms/unit cell equal to 8(16); a combinatorial argument gives a total of $300$ possible stoichiometries.
For each pressure, we generated a total of 1800 structures, which gives an average of $6$ structures/composition. 
We would like to remind that this is only an exploratory run,
while more accurate runs were used to inspect the most promising compositions.~\cite{CD_note1}

Despite the apparently coarse sampling, our preliminary search identified correctly all known Li-B-H phases, both along the boundary lines and in the 
middle of the phase diagram. Only Li$2$B$_{12}$H$_{12}$, which is an important intermediate product of the hydrogenation process of LiBH$_4$,  has been added by hand, because the unit cell at ambient pressure is larger than the maximum
number of atoms employed for our search.
At ambient pressure, we reproduce the phase diagram and energetics of previous works; at P=200 GPa, there are no 
literature data for ternary phases, 
but we reproduce known results for the Li-H, B-H and Li-B systems.~\cite{zurek2009little,kolmogorov_lib_PRB2006,PhysRevLett.110.165504}

Our previous experience on binary systems taught us that the energies and 
structures from initial
coarse sampling runs need to further be refined to obtain a correct ranking of structures and compositions.\cite{kokail_PhysRevB.94.060502}
For this reason, after the initial scan, we 
focused on to two specific Li:B:H lines, shown in Fig.~\ref{fig:TPD}.~\cite{CD_NOTE2}
These are the 1:1:$x$ line, that contains compounds with chemical formula LiBH$_x$, including LiBH$_4$,
and the 2:1:$x$ one, where we found a highly symmetric metallic structure with chemical formula Li$_2$BH$_6$.
For these two lines, we ran additional crystal structure prediction runs with
tighter settings; the same was done for boundary lines, and for 
 Li$_2$B$_{12}$H$_{12}$.

The two enthalpy ($\Delta H$) vs. composition ($x$) binary
convex hulls are shown in the two lower panels of Fig.~\ref{fig:TPD}.

Similarly to what observed in binary hydrides, pressures in the Megabar range
stabilize several compositions which are metastable at ambient pressure.
In particular, along the 1:1:$x$ line LiBH, LiBH$_2$ and LiBH$_6$, besides the ground-state LiBH$_4$, lie close to the hull, while for the 2:1:$x$ 
line there are several compositions close to the hull.
Note that compositions on the binary hull are stable with respect to the decomposition into the end members of the line (LiB + H, and Li$_2$B + H); however, in a ternary system other decompositions are also possible. Although computing
all possible paths would be prohibitive, 
we recomputed the enthalpy of formation of all compounds on the binary hulls 
also with respect to boundary phases; taking this effect into account, a few phases on the binary hull turned out to be metastable.
These are shown as (blue) circles in Fig.~\ref{fig:TPD}, while genuinely ground state structures are shown as (red) squares.
In the following, we will discuss the crystal and electronic structure of the most interesting
compositions, with the aim of identifying potential high-\tc\; superconductors. 

\begin{figure}[t]
\includegraphics[width=1.0\columnwidth,angle=0]{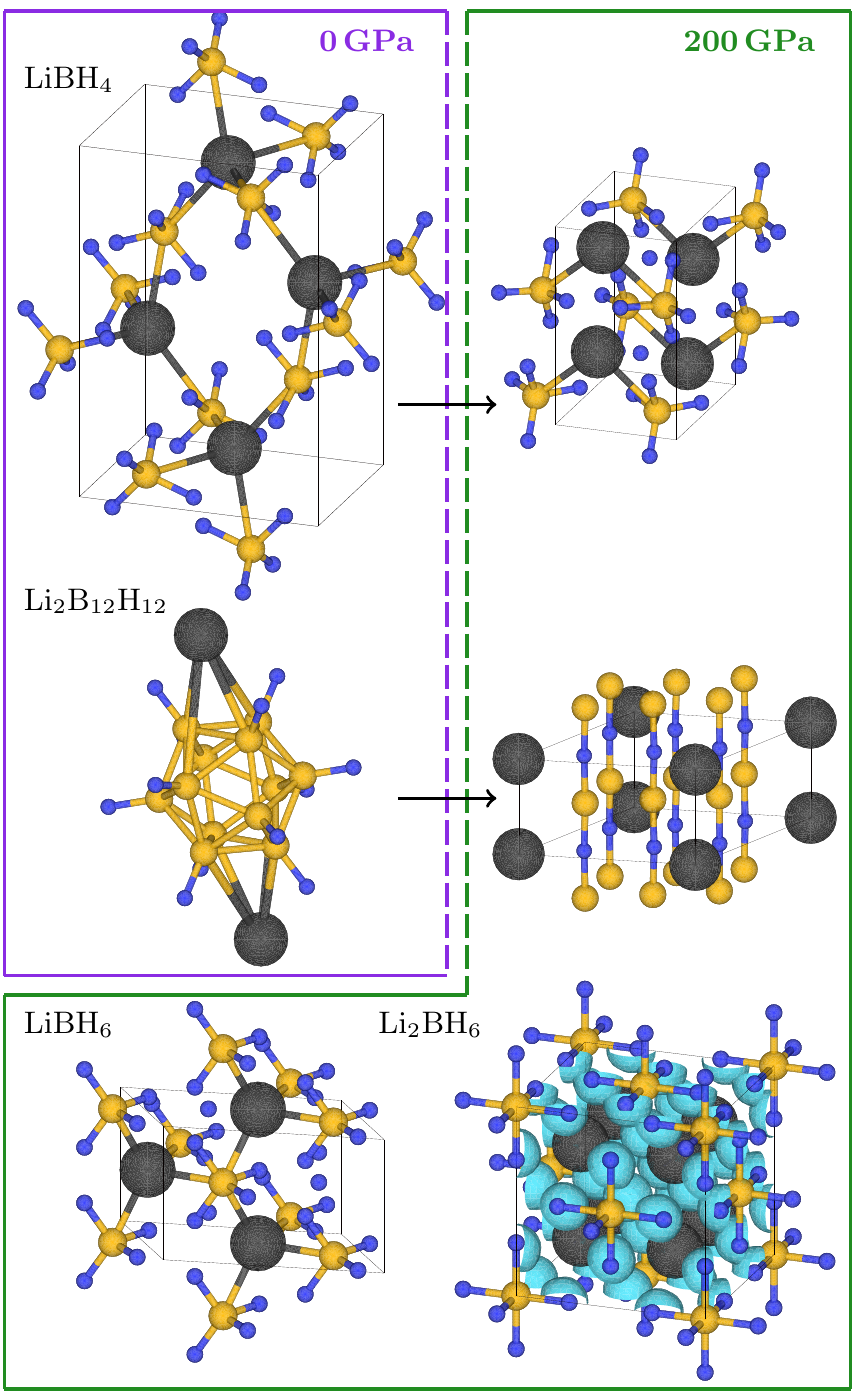}
\caption{Crystal structure of selected Li-B-H
phases at 0 and 200 GPa identified in this work. For Li$_2$BH$_6$,
we also plot the 0.7 ELF isocontour.
}
\label{fig:crystal}
\end{figure}

We start from the ground-state LiBH$_4$, shown in the upper panel of Fig.~\ref{fig:crystal}.
 For this stoichiometry,
we ran  evolutionary structure prediction runs at fixed compositions for 
0, 100, 200 and 300$\,$ GPa
with 2,3 and 4 formula units per unit cell. 
At all pressures, we found as most stable a $Pmma$ structure,
in which (BH$_4$)$^{-}$ tetrahedra are intercalated with $Li^+$ ions.
At ambient pressure, the structure is very open, and the BH$_4$
tetrahedra can orient freely in the unit cell. Pressure leads to
a more close-packed arrangement, in which the BH$_4$ tetrahedra
only acquire two possible orientations around the Li atoms.
The high-pressure structure shown in the figure 
is stable at least up to 300 GPa, where it is still insulating.
Thus, LiBH$_4$ cannot support 
high-\tc\, conventional superconductivity as in \sh,
but other phases on the phase diagrams are strong candidates.

An obvious candidate, due to its high hydrogen content, 
is Li$_2$B$_{12}$H$_{12}$. At ambient pressure, this compound 
crystallizes in an open structure of B-H icosahedra, 
intercalated with lithium atoms. Icosahedra are found in 
$\alpha$-boron and in several B-rich phases, 
including superconducting dodecaborides, such as ZrB$_{12}$.\cite{Matthias_B12}
However, at ambient pressure Li$_2$B$_{12}$H$_{12}$ is insulating,
and hence cannot superconduct.
At higher pressures, the eicosahedral environment is destabilized,
and Li$_2$B$_{12}$H$_{12}$ acquires a completely different structures, characterized by unidimensional B-H chains,~\cite{PhysRevLett.110.165504}
intercalated by lithium. 
This phase is however metastable (by $200$ meV/atom)
with respect to elemental decomposition, and we will not consider it further
in our study.

Other compounds which have been often discussed in the hydrogenation
and rehydrogenation reactions of LiBH$_4$ are those that lie
along the 1:1:$x$ Li:B:H line. The bottom left panel of Fig.~\ref{fig:crystal}
shows the high-pressure crystal structure of LiBH$_6$. 
The high-pressure stabilization of a hydrogen-rich phase of LiBH$_4$ 
could be the analogue of the reaction SH$_2$ + H$_2$ $\rightarrow$ \sh\;
that led to the discovery of the first high-pressure conventional 
superconductor.
However, Fig.~\ref{fig:crystal} shows that there is an important difference
between \sh\; and LiBH$_6$. In \sh\; 
a pressure of 200 GPa is sufficient to break the molecular bonds of SH$_2$ and H$_2$ and stabilize three new directional, covalent
bonds between S and H. In LiBH$_6$ one can still recognize a close-packed
LiBH$_4$ lattice, and molecular hydrogen intercalated in-between. This structure
should thus rather be described as LiBH$_4$ + H$_2$ than LiBH$_6$. 
Not surprisingly, this structure is insulating.

Our evolutionary runs allowed us to identify at least one hydrogen-rich
phase in which the (BH$_4$) tetrahedral environment is destabilized,
and molecular hydrogen is incorporated into the boron lattice.
This is the Li$_2$BH$_6$ structure shown in the bottom right panel
of Fig.~\ref{fig:crystal}. Here, boron and hydrogen form octahedra,
and lithium sits in-between.
BH$_6$ octahedra are not common in nature, but an AlH$_6$
octahedral motif is common in alanates.~\cite{Peles_PRB_2004,marques_PRL_2013}
For borohydrides this motif, which is stabilized by 
$e_g$ ($d$) electrons, has never been 
observed at ambient pressure, and we consider our finding an
example of high-pressure forbidden chemistry; we will come
back to this point in the following.

 Although unusual in structure and composition,
according to our calculations Li$_2$BH$_6$ remains thermodynamically
stable with respect to decompositions towards all phases
on the ternary Gibbs diagram down to 100 GPa.
Given the accuracy of our predictions in all other cases for which
we had access to experimental data, we believe that this is
a strong indication that Li$_2$BH$_6$ could be synthesized in experiments. 
In Fig.~\ref{fig:crystal}, superimposed to the crystal structure 
of Li$_2$BH$_6$, we show the 0.7 isocontour of the electronic localization
function (ELF); the plot shows that most of the charge resides along the
BH$_6$ bonds. Combined with the fact that Li$_2$BH$_6$ is metallic,
this makes it a very strong candidate for high-\tc\; 
conventional superconductivity. Indeed, as we will show, our electronic structure calculations confirm this hypothesis.

\begin{figure}[t]
\includegraphics[width=1.0\columnwidth,angle=0]{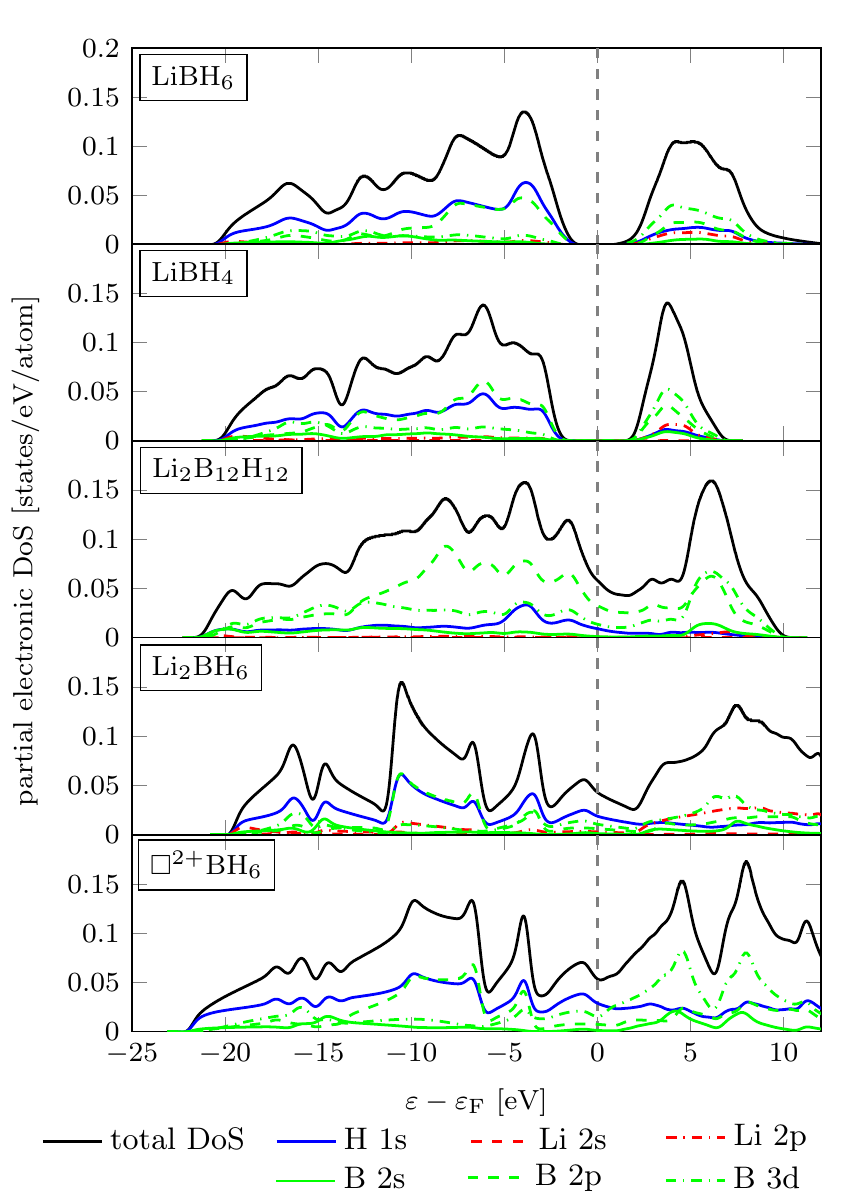}
\caption{Partial Density of States (pDOS's) of the most representative
phases analysed in this work. All DOS's have been calculated at 200 GPa.
From top to bottom: LiBH$_6$, LiBH$_4$,
Li$_2$B$_{12}$H$_{12}$, Li$_2$BH$_6$, and a hypothetical compound in 
which Li has been replaced by a uniform, positive background of 
charge -- ($\square^{+2}$BH$_6$).}
\label{fig:pdos}
\end{figure}

Fig.~\ref{fig:pdos} shows the partial Densities of States (DOS) 
of the most relevant ternary Li:B:H phases
in this work, calculated at 200 GPa.
The first two panels show LiBH$_4$ and LiBH$_6$; in both compounds
a large gap ($\Delta \sim 3 eV$) separates
bonding and antibonding states derived from the hybridization 
of B sp$^3$ states with hydrogen. This makes the BH$_4$ 
environment extremely stable; in fact, in LiBH$_6$
the two excess hydrogens do not bind to boron, but
remain in molecular form, and arrange in the interstitials of the structure;
the relative electronic states
form an additional peak near the top of the valence band.
Other structures along the 1:1:$x$ line (not shown) are also insulating 
for similar reasons at this pressure.
Li$_2$B$_2$H$_{12}$, shown immediately below,
is a good metal, but metastable.

The two bottom panels show 
Li$_2$B$_6$H$_6$, which is the most promising
candidate for superconductivity identified in this work,
and a hypothetical compound in which lithium is replaced
by a uniform background of charge ($\square^{+2}$BH$_6$).
The strong similarity between the two DOS's in the valence region indicates
that the main role of lithium in this structure is to donate charge
to the boron-hydrogen octahedra, while its contribution to the
bonding is only marginal.

We can thus try to understand the electronic structure in terms of
the BH$_6$ cluster alone; the states at the Fermi level
result from the hybridization of B $d$ $e_g$ states with hydrogen;
the two other structures centered at $\sim -8$ eV and $\sim -15$
eV correspond to B $s$ and $p$ states.
It has been argued that 
the octahedral environment is not seen in borohydrides, because the
gap between $d$ and $p$ states is too large compared to other hydrides
of the third group.
In these compounds, the $XH_6$ environment is stable already at 
ambient pressure, where the bandwidth is much smaller. 
$s$, $p$ and $e_g$ states cause clear gaps in the electronic spectrum.
Octahedral hydrides typically host 12 valence electrons,
corresponding to a complete filling of $s$, $p$ and $e_g$ shells.  
On the other hand,  according to our calculations, 
the Li$_2$BH$_6$ phase, which has only 11 valence electrons, 
is thermodynamically
stable down to 100 GPa, where it remains metallic.
We believe that the reason why this unusual phase can occur at high
pressures is that the boron-hydrogen bandwidth is
large enough to overcome the intrinsic gaps in the boron spectrum,
giving rise to a metallic DOS, allowing a wider range of dopings.
 In Li$2$BH$_6$, the Fermi level
sits in a shallow region of this continuum, where $N \sim 0.2$ st/eV f.u.

In order to estimate the actual superconducting characteristics
of Li$_2$BH$_6$, we 
performed linear response calculations~\cite{Baroni_RMP2001} 
of its electron-phonon properties,
and estimated the critical temperature through the Mc-Millan Allen-Dynes formula~\cite{McMillanTC,AllenDynes_PRB1975}:
\begin{align} \label{eq:tc}
T_\text{c} = \frac{\omega_\text{log}}{1.2 k_\text{B}} \exp \left[ -\frac{1.04 (1+\lambda)}{\lambda-\mu^{\star}(1+0.62 \lambda)} \right],
\end{align}

\begin{figure}[t]
\includegraphics[width=1.0\columnwidth,angle=0]{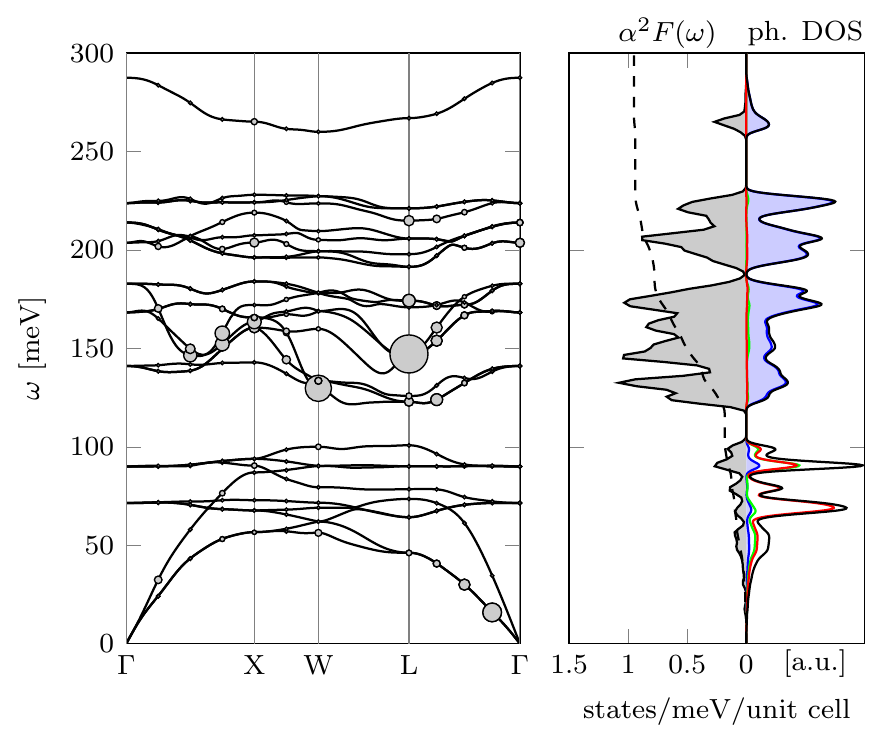}
\includegraphics[width=1.0\columnwidth,angle=0]{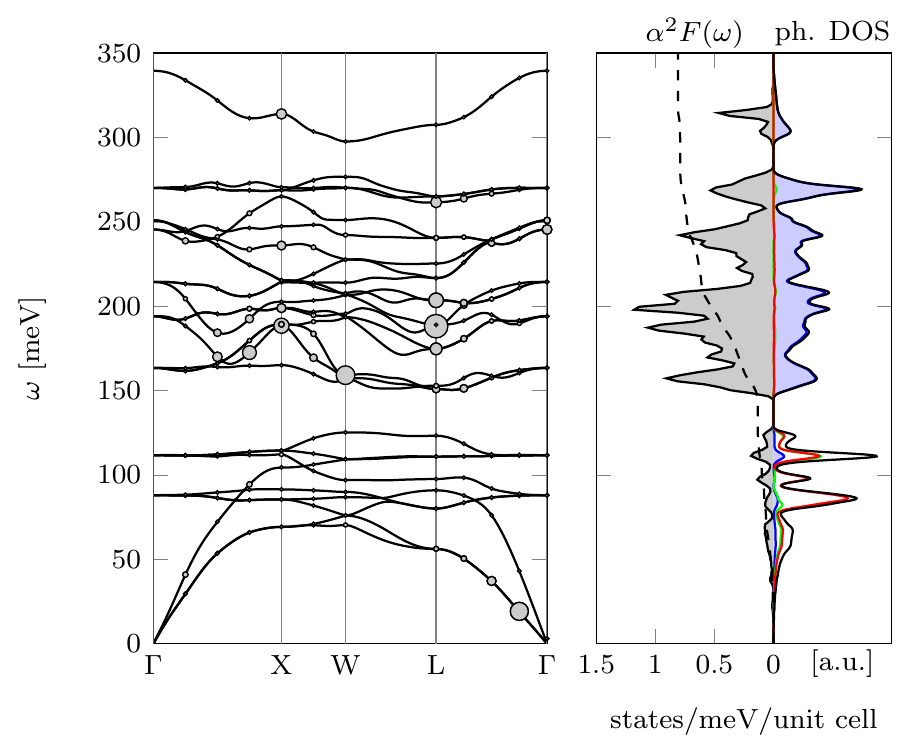}
\caption{Phonon dispersions (left), densities of states 
and $ep$ spectral function (right) of Li$_2$BH$_6$ at 100 (top) and 200 (bottom) GPa. 
The size of the circles in the phonon dispersion plot is proportional
to the partial $ep$ coupling constant for a given mode. The colors
in the DOS plot indicate partial contributions; the color coding is the same
as in Fig.~\ref{fig:pdos}.}
\label{fig:alpha}
\end{figure}

The phonon dispersions, decorated with circles whose 
size is proportional to the partial $ep$ coupling of each branch,
are shown in the left panels of Fig.~\ref{fig:alpha};
the right panels shows the partial Phonon DOS and the
$ep$ (Eliashberg) spectral function $\alpha^2 F(\omega)$,
which describes how the $ep$  coupling is distributed
on phonon modes with energy $\omega$. The top and bottom
panels refer to P=100 and P=200 GPa, respectively.
The parameters $\lambda$ ($ep$ coupling constant) and \omlog (logaritmically averaged phonon frequency) in
Eq.~\ref{eq:tc} can be obtained from $\alpha^2 F(\omega)$ as:
 $\lambda=2\int d\omega \frac{\alpha^2F(\omega)}{\omega}$ and 
$\omega_\text{log} = \exp \left[ \frac{2}{\lambda} \int \frac{d\omega}{\omega} \alpha^2 F(\omega)\ln(\omega)\right]$; $\mu^{\star}$ is the Coulomb pseudopotential, renormalized to include retardation effects due to the large disparity between the electron and phonon energies.

For P=100 and 200 GPa, we obtain \omlog=1551 and 1940 $K$
and $\lambda$=0.94, 0.76, respectively.  The corresponding \tc's, estimated from Eq.~\ref{eq:tc} with $\mu^*=0.1$ are 98 and 81 K, comparable to
those of \ph.
Comparing the $\alpha^2 F(\omega)$ with the partial phonon DOS next to it,
it is clear that the most substantial contribution to the coupling comes 
from H modes at intermediate frequencies, while octahedral vibrations 
(above 300 meV), play a very marginal role. 
The phonon spectrum shifts almost rigidly by $\sim 50$ meV 
going from 100 to 200 GPa, causing a similar increase in \omlog.
On the other hand, the electronic properties worsen with pressure,
since the DOS at the Fermi level decreases by 20\%, causing a similar
decrease in the total $ep$ coupling constant.
The factor $\eta=\lambda/N(E_F)$, which is 
a measure of the lattice contribution to the $ep$ coupling,
is almost constant $\eta \simeq 4.2$ eV f.u., and comparable with
\sh\; and \ph, where $\eta$ is 3.6 and 3.8, 
respectively.~\cite{Flores_PH3_PRBR2016}
Fig.~\ref{fig:pdos} shows that the Fermi level in Li$_2$BH$_6$
sits in a shallow region of the DOS, which is weakly affected by pressure;
this explains the weak dependence of \tc\; on $P$. 

While this weak dependence
implies that \tc\; cannot 
be effectively boosted by pressure, as in \ph and \sh,
it also implies that superconductivity survives with remarkable \tc's down
to pressures which are twice smaller than in \sh.
Furthermore, the fact that the atoms that contribute to charge
doping and covalent bonding are different (lithium and boron, respectively),
offers a simple route to improve the superconducting properties of Li$_2$BH$_6$.
Partially replacing lithium 
with alkaline earths or vacancies would allow to easily tune the
doping level, and hence the value of the DOS, without
affecting the stiff boron-hydrogen sublattice responsible for the
large $ep$ coupling. Doping on the Li site in ternary
hydrides is routinely achieved in hydrogen storage applications,
and is most likely much easier to obtain also at high pressures
than the iso- or heterovalent
substitutions proposed by several 
authors for covalent hydrides.~\cite{Heil-Boeri_PRB2015,Ge_PRB_2016,Flores_H2O}
On the other hand, substitutions at the B site could be used
to tune other intrinsic properties, such as $ep$ matrix elements
or metallization pressures.

In conclusion, in this work we have studied from first-principles 
the high-pressure superconducting  phase diagram of lithium-boron-hydrogen, a prototypical ternary system employed for
hydrogen storage applications. 
Besides the well-known boundary phases, we have identified several new compositions which are stabilized by high pressures. We have shown that 
neither the ground-state LiBH$_4$, nor any of its direct hydrogenation or 
dehydrogenation products 
is a viable candidate for high-\tc~ superconductivity, but we have
identified at least one ternary phase, Li$_2$BH$_6$, which 
exhibits superconducting
properties comparable to those of the best binary hydrides.
The Li$_2$BH$_6$ composition is not stable at ambient pressure, but 
according to our calculations it should become thermodynamically stable
for P > 100 GPa. Similarly to \sh, which is a hydrogen-rich phase of sulfur hydride, 
in which the original molecular bonds are broken and new, directional bonds are formed under pressure,
Li$_2$BH$_6$ exhibits a highly symmetric structure in which the original BH$_4$ tetrahedra
that are characteristic of boronhydrides rearrange to form BH$_6$ octahedra, with covalent B-H bonds. These determine the valence band structure,
while lithium mainly acts as a charge reservoir. The fact that 
two different atoms govern the bonding and the charge doping should
allow to tune the \tc\;  more easily than in binary hydrides.
Our work demonstrates that ternary hydrides can exhibit high-\tc\ superconductivity and is a first step towards the optimization of superconducting properties in high-pressure hydrides using chemical methods.

\begin{acknowledgments}
 The authors acknowledge computational resources from the dCluster of the Graz University of Technology and  the VSC3 of the Vienna University of Technology,
and support through the FWF project P 30269-N36 ({\em Superhydra}).
\end{acknowledgments}

\bibliographystyle{apsrev4-1}


\begin{thebibliography}{80}%
\makeatletter
\providecommand \@ifxundefined [1]{%
 \@ifx{#1\undefined}
}%
\providecommand \@ifnum [1]{%
 \ifnum #1\expandafter \@firstoftwo
 \else \expandafter \@secondoftwo
 \fi
}%
\providecommand \@ifx [1]{%
 \ifx #1\expandafter \@firstoftwo
 \else \expandafter \@secondoftwo
 \fi
}%
\providecommand \natexlab [1]{#1}%
\providecommand \enquote  [1]{``#1''}%
\providecommand \bibnamefont  [1]{#1}%
\providecommand \bibfnamefont [1]{#1}%
\providecommand \citenamefont [1]{#1}%
\providecommand \href@noop [0]{\@secondoftwo}%
\providecommand \href [0]{\begingroup \@sanitize@url \@href}%
\providecommand \@href[1]{\@@startlink{#1}\@@href}%
\providecommand \@@href[1]{\endgroup#1\@@endlink}%
\providecommand \@sanitize@url [0]{\catcode `\\12\catcode `\$12\catcode
  `\&12\catcode `\#12\catcode `\^12\catcode `\_12\catcode `\%12\relax}%
\providecommand \@@startlink[1]{}%
\providecommand \@@endlink[0]{}%
\providecommand \url  [0]{\begingroup\@sanitize@url \@url }%
\providecommand \@url [1]{\endgroup\@href {#1}{\urlprefix }}%
\providecommand \urlprefix  [0]{URL }%
\providecommand \Eprint [0]{\href }%
\providecommand \doibase [0]{http://dx.doi.org/}%
\providecommand \selectlanguage [0]{\@gobble}%
\providecommand \bibinfo  [0]{\@secondoftwo}%
\providecommand \bibfield  [0]{\@secondoftwo}%
\providecommand \translation [1]{[#1]}%
\providecommand \BibitemOpen [0]{}%
\providecommand \bibitemStop [0]{}%
\providecommand \bibitemNoStop [0]{.\EOS\space}%
\providecommand \EOS [0]{\spacefactor3000\relax}%
\providecommand \BibitemShut  [1]{\csname bibitem#1\endcsname}%
\let\auto@bib@innerbib\@empty
\bibitem [{\citenamefont {Ashcroft}(1968)}]{PhysRevLett.21.1748}%
  \BibitemOpen
  \bibfield  {author} {\bibinfo {author} {\bibfnamefont {N.~W.}\ \bibnamefont
  {Ashcroft}},\ }\href {\doibase 10.1103/PhysRevLett.21.1748} {\bibfield
  {journal} {\bibinfo  {journal} {Phys. Rev. Lett.}\ }\textbf {\bibinfo
  {volume} {21}},\ \bibinfo {pages} {1748} (\bibinfo {year}
  {1968})}\BibitemShut {NoStop}%
\bibitem [{\citenamefont {Wigner}\ and\ \citenamefont
  {Huntington}(1935)}]{Wigner_JCP1935}%
  \BibitemOpen
  \bibfield  {author} {\bibinfo {author} {\bibfnamefont {E.}~\bibnamefont
  {Wigner}}\ and\ \bibinfo {author} {\bibfnamefont {H.~B.}\ \bibnamefont
  {Huntington}},\ }\href {\doibase http://dx.doi.org/10.1063/1.1749590}
  {\bibfield  {journal} {\bibinfo  {journal} {The Journal of Chemical Physics}\
  }\textbf {\bibinfo {volume} {3}},\ \bibinfo {pages} {764} (\bibinfo {year}
  {1935})}\BibitemShut {NoStop}%
\bibitem [{\citenamefont {Cudazzo}\ \emph {et~al.}(2008)\citenamefont
  {Cudazzo}, \citenamefont {Profeta}, \citenamefont {Sanna}, \citenamefont
  {Floris}, \citenamefont {Continenza}, \citenamefont {Massidda},\ and\
  \citenamefont {Gross}}]{PhysRevLett.100.257001}%
  \BibitemOpen
  \bibfield  {author} {\bibinfo {author} {\bibfnamefont {P.}~\bibnamefont
  {Cudazzo}}, \bibinfo {author} {\bibfnamefont {G.}~\bibnamefont {Profeta}},
  \bibinfo {author} {\bibfnamefont {A.}~\bibnamefont {Sanna}}, \bibinfo
  {author} {\bibfnamefont {A.}~\bibnamefont {Floris}}, \bibinfo {author}
  {\bibfnamefont {A.}~\bibnamefont {Continenza}}, \bibinfo {author}
  {\bibfnamefont {S.}~\bibnamefont {Massidda}}, \ and\ \bibinfo {author}
  {\bibfnamefont {E.~K.~U.}\ \bibnamefont {Gross}},\ }\href {\doibase
  10.1103/PhysRevLett.100.257001} {\bibfield  {journal} {\bibinfo  {journal}
  {Phys. Rev. Lett.}\ }\textbf {\bibinfo {volume} {100}},\ \bibinfo {pages}
  {257001} (\bibinfo {year} {2008})}\BibitemShut {NoStop}%
\bibitem [{\citenamefont {McMahon}\ and\ \citenamefont
  {Ceperley}(2011)}]{PhysRevB.84.144515}%
  \BibitemOpen
  \bibfield  {author} {\bibinfo {author} {\bibfnamefont {J.~M.}\ \bibnamefont
  {McMahon}}\ and\ \bibinfo {author} {\bibfnamefont {D.~M.}\ \bibnamefont
  {Ceperley}},\ }\href {\doibase 10.1103/PhysRevB.84.144515} {\bibfield
  {journal} {\bibinfo  {journal} {Phys. Rev. B}\ }\textbf {\bibinfo {volume}
  {84}},\ \bibinfo {pages} {144515} (\bibinfo {year} {2011})}\BibitemShut
  {NoStop}%
\bibitem [{\citenamefont {Borinaga}\ \emph {et~al.}(2016)\citenamefont
  {Borinaga}, \citenamefont {Errea}, \citenamefont {Calandra}, \citenamefont
  {Mauri},\ and\ \citenamefont {Bergara}}]{PhysRevB.93.174308}%
  \BibitemOpen
  \bibfield  {author} {\bibinfo {author} {\bibfnamefont {M.}~\bibnamefont
  {Borinaga}}, \bibinfo {author} {\bibfnamefont {I.}~\bibnamefont {Errea}},
  \bibinfo {author} {\bibfnamefont {M.}~\bibnamefont {Calandra}}, \bibinfo
  {author} {\bibfnamefont {F.}~\bibnamefont {Mauri}}, \ and\ \bibinfo {author}
  {\bibfnamefont {A.}~\bibnamefont {Bergara}},\ }\href {\doibase
  10.1103/PhysRevB.93.174308} {\bibfield  {journal} {\bibinfo  {journal} {Phys.
  Rev. B}\ }\textbf {\bibinfo {volume} {93}},\ \bibinfo {pages} {174308}
  (\bibinfo {year} {2016})}\BibitemShut {NoStop}%
\bibitem [{\citenamefont {Szczesniak}\ and\ \citenamefont
  {Jarosik}(2009)}]{Szniak20092053}%
  \BibitemOpen
  \bibfield  {author} {\bibinfo {author} {\bibfnamefont {R.}~\bibnamefont
  {Szczesniak}}\ and\ \bibinfo {author} {\bibfnamefont {M.}~\bibnamefont
  {Jarosik}},\ }\href {\doibase http://dx.doi.org/10.1016/j.ssc.2009.08.019}
  {\bibfield  {journal} {\bibinfo  {journal} {Solid State Communications}\
  }\textbf {\bibinfo {volume} {149}},\ \bibinfo {pages} {2053 } (\bibinfo
  {year} {2009})}\BibitemShut {NoStop}%
\bibitem [{\citenamefont {Eremets}\ \emph {et~al.}(2016)\citenamefont
  {Eremets}, \citenamefont {Troyan},\ and\ \citenamefont
  {Drozdov}}]{eremets2016low}%
  \BibitemOpen
  \bibfield  {author} {\bibinfo {author} {\bibfnamefont {M.}~\bibnamefont
  {Eremets}}, \bibinfo {author} {\bibfnamefont {I.}~\bibnamefont {Troyan}}, \
  and\ \bibinfo {author} {\bibfnamefont {A.}~\bibnamefont {Drozdov}},\ }\href
  {https://arxiv.org/abs/1601.04479} {\bibfield  {journal} {\bibinfo  {journal}
  {arXiv preprint arXiv:1601.04479}\ } (\bibinfo {year} {2016})}\BibitemShut
  {NoStop}%
\bibitem [{\citenamefont {Ross}\ and\ \citenamefont
  {Eugene}(2016)}]{nat0028-0836}%
  \BibitemOpen
  \bibfield  {author} {\bibinfo {author} {\bibfnamefont {P.~D.-S. P. T.~H.}\
  \bibnamefont {Ross}}\ and\ \bibinfo {author} {\bibfnamefont {E.~G.}\
  \bibnamefont {Eugene}},\ }\href {\doibase
  http://dx.doi.org/10.1038/nature16164 10.1038/nature16164} {\bibfield
  {journal} {\bibinfo  {journal} {Nature}\ }\textbf {\bibinfo {volume} {529}},\
  \bibinfo {pages} {63} (\bibinfo {year} {2016})}\BibitemShut {NoStop}%
\bibitem [{\citenamefont {Babaev}\ \emph {et~al.}(2004)\citenamefont {Babaev},
  \citenamefont {Sudbo},\ and\ \citenamefont {Ashcroft}}]{babaev}%
  \BibitemOpen
  \bibfield  {author} {\bibinfo {author} {\bibfnamefont {E.}~\bibnamefont
  {Babaev}}, \bibinfo {author} {\bibfnamefont {A.}~\bibnamefont {Sudbo}}, \
  and\ \bibinfo {author} {\bibfnamefont {N.~W.}\ \bibnamefont {Ashcroft}},\
  }\href {\doibase http://dx.doi.org/10.1038/nature02910} {\bibfield  {journal}
  {\bibinfo  {journal} {Nature}\ }\textbf {\bibinfo {volume} {431}},\ \bibinfo
  {pages} {666} (\bibinfo {year} {2004})},\ \bibinfo {note}
  {10.1038/nature02910}\BibitemShut {NoStop}%
\bibitem [{\citenamefont {Dias}\ and\ \citenamefont {Silvera}(2017)}]{Dias715}%
  \BibitemOpen
  \bibfield  {author} {\bibinfo {author} {\bibfnamefont {R.~P.}\ \bibnamefont
  {Dias}}\ and\ \bibinfo {author} {\bibfnamefont {I.~F.}\ \bibnamefont
  {Silvera}},\ }\href {\doibase 10.1126/science.aal1579} {\bibfield  {journal}
  {\bibinfo  {journal} {Science}\ }\textbf {\bibinfo {volume} {355}},\ \bibinfo
  {pages} {715} (\bibinfo {year} {2017})}\BibitemShut {NoStop}%
\bibitem [{\citenamefont {Goncharov}\ and\ \citenamefont
  {Struzhkin}(2017)}]{goncharov2017comment}%
  \BibitemOpen
  \bibfield  {author} {\bibinfo {author} {\bibfnamefont {A.~F.}\ \bibnamefont
  {Goncharov}}\ and\ \bibinfo {author} {\bibfnamefont {V.~V.}\ \bibnamefont
  {Struzhkin}},\ }\href {https://arxiv.org/abs/1702.04246} {\bibfield
  {journal} {\bibinfo  {journal} {arXiv preprint arXiv:1702.04246}\ } (\bibinfo
  {year} {2017})}\BibitemShut {NoStop}%
\bibitem [{\citenamefont {Eremets}\ and\ \citenamefont
  {Drozdov}(2017)}]{eremets_comment}%
  \BibitemOpen
  \bibfield  {author} {\bibinfo {author} {\bibfnamefont {M.}~\bibnamefont
  {Eremets}}\ and\ \bibinfo {author} {\bibfnamefont {A.~P.}\ \bibnamefont
  {Drozdov}},\ }\href {arXiv:1702.05125} {\bibfield  {journal} {\bibinfo
  {journal} {airxiv-condmat/1702.05125}\ } (\bibinfo {year}
  {2017})}\BibitemShut {NoStop}%
\bibitem [{\citenamefont {Ashcroft}(2004)}]{PhysRevLett.92.187002}%
  \BibitemOpen
  \bibfield  {author} {\bibinfo {author} {\bibfnamefont {N.~W.}\ \bibnamefont
  {Ashcroft}},\ }\href {\doibase 10.1103/PhysRevLett.92.187002} {\bibfield
  {journal} {\bibinfo  {journal} {Phys. Rev. Lett.}\ }\textbf {\bibinfo
  {volume} {92}},\ \bibinfo {pages} {187002} (\bibinfo {year}
  {2004})}\BibitemShut {NoStop}%
\bibitem [{\citenamefont {Yao}\ and\ \citenamefont
  {Klug}(2010)}]{yao2010silane}%
  \BibitemOpen
  \bibfield  {author} {\bibinfo {author} {\bibfnamefont {Y.}~\bibnamefont
  {Yao}}\ and\ \bibinfo {author} {\bibfnamefont {D.~D.}\ \bibnamefont {Klug}},\
  }\href {http://www.pnas.org/content/107/49/20893.short} {\bibfield  {journal}
  {\bibinfo  {journal} {Proceedings of the National Academy of Sciences}\
  }\textbf {\bibinfo {volume} {107}},\ \bibinfo {pages} {20893} (\bibinfo
  {year} {2010})}\BibitemShut {NoStop}%
\bibitem [{\citenamefont {Feng}\ \emph {et~al.}(2006)\citenamefont {Feng},
  \citenamefont {Grochala}, \citenamefont {Jaro\ifmmode~\acute{n}\else
  \'{n}\fi{}}, \citenamefont {Hoffmann}, \citenamefont {Bergara},\ and\
  \citenamefont {Ashcroft}}]{PhysRevLett.96.017006}%
  \BibitemOpen
  \bibfield  {author} {\bibinfo {author} {\bibfnamefont {J.}~\bibnamefont
  {Feng}}, \bibinfo {author} {\bibfnamefont {W.}~\bibnamefont {Grochala}},
  \bibinfo {author} {\bibfnamefont {T.}~\bibnamefont
  {Jaro\ifmmode~\acute{n}\else \'{n}\fi{}}}, \bibinfo {author} {\bibfnamefont
  {R.}~\bibnamefont {Hoffmann}}, \bibinfo {author} {\bibfnamefont
  {A.}~\bibnamefont {Bergara}}, \ and\ \bibinfo {author} {\bibfnamefont
  {N.~W.}\ \bibnamefont {Ashcroft}},\ }\href {\doibase
  10.1103/PhysRevLett.96.017006} {\bibfield  {journal} {\bibinfo  {journal}
  {Phys. Rev. Lett.}\ }\textbf {\bibinfo {volume} {96}},\ \bibinfo {pages}
  {017006} (\bibinfo {year} {2006})}\BibitemShut {NoStop}%
\bibitem [{\citenamefont {Drozdov}\ \emph
  {et~al.}(2015{\natexlab{a}})\citenamefont {Drozdov}, \citenamefont {I.},
  \citenamefont {Troyan}, \citenamefont {Ksenofontov},\ and\ \citenamefont
  {Shylin}}]{DrozdovEremets_Nature2015}%
  \BibitemOpen
  \bibfield  {author} {\bibinfo {author} {\bibfnamefont {A.~P.}\ \bibnamefont
  {Drozdov}}, \bibinfo {author} {\bibfnamefont {M.~E.~M.}\ \bibnamefont {I.}},
  \bibinfo {author} {\bibfnamefont {I.}~\bibnamefont {Troyan}}, \bibinfo
  {author} {\bibfnamefont {V.}~\bibnamefont {Ksenofontov}}, \ and\ \bibinfo
  {author} {\bibfnamefont {S.}~\bibnamefont {Shylin}},\ }\href {\doibase
  http://dx.doi.org/10.1038/nature14964 10.1038/nature14964} {\bibfield
  {journal} {\bibinfo  {journal} {Nature}\ }\textbf {\bibinfo {volume} {525}},\
  \bibinfo {pages} {73} (\bibinfo {year} {2015}{\natexlab{a}})}\BibitemShut
  {NoStop}%
\bibitem [{\citenamefont {Troyan}\ \emph {et~al.}(2016)\citenamefont {Troyan},
  \citenamefont {Gavriliuk}, \citenamefont {R{\"u}ffer}, \citenamefont
  {Chumakov}, \citenamefont {Mironovich}, \citenamefont {Lyubutin},
  \citenamefont {Perekalin}, \citenamefont {Drozdov},\ and\ \citenamefont
  {Eremets}}]{Troyan1303}%
  \BibitemOpen
  \bibfield  {author} {\bibinfo {author} {\bibfnamefont {I.}~\bibnamefont
  {Troyan}}, \bibinfo {author} {\bibfnamefont {A.}~\bibnamefont {Gavriliuk}},
  \bibinfo {author} {\bibfnamefont {R.}~\bibnamefont {R{\"u}ffer}}, \bibinfo
  {author} {\bibfnamefont {A.}~\bibnamefont {Chumakov}}, \bibinfo {author}
  {\bibfnamefont {A.}~\bibnamefont {Mironovich}}, \bibinfo {author}
  {\bibfnamefont {I.}~\bibnamefont {Lyubutin}}, \bibinfo {author}
  {\bibfnamefont {D.}~\bibnamefont {Perekalin}}, \bibinfo {author}
  {\bibfnamefont {A.~P.}\ \bibnamefont {Drozdov}}, \ and\ \bibinfo {author}
  {\bibfnamefont {M.~I.}\ \bibnamefont {Eremets}},\ }\href {\doibase
  10.1126/science.aac8176} {\bibfield  {journal} {\bibinfo  {journal}
  {Science}\ }\textbf {\bibinfo {volume} {351}},\ \bibinfo {pages} {1303}
  (\bibinfo {year} {2016})}\BibitemShut {NoStop}%
\bibitem [{\citenamefont {Duan}\ \emph {et~al.}(2014)\citenamefont {Duan},
  \citenamefont {Liu}, \citenamefont {Tian}, \citenamefont {Li}, \citenamefont
  {Huang}, \citenamefont {Zhao}, \citenamefont {Yu}, \citenamefont {Liu},
  \citenamefont {Tian},\ and\ \citenamefont {Cui}}]{Duan_SciRep2014}%
  \BibitemOpen
  \bibfield  {author} {\bibinfo {author} {\bibfnamefont {D.}~\bibnamefont
  {Duan}}, \bibinfo {author} {\bibfnamefont {Y.}~\bibnamefont {Liu}}, \bibinfo
  {author} {\bibfnamefont {F.}~\bibnamefont {Tian}}, \bibinfo {author}
  {\bibfnamefont {D.}~\bibnamefont {Li}}, \bibinfo {author} {\bibfnamefont
  {X.}~\bibnamefont {Huang}}, \bibinfo {author} {\bibfnamefont
  {Z.}~\bibnamefont {Zhao}}, \bibinfo {author} {\bibfnamefont {H.}~\bibnamefont
  {Yu}}, \bibinfo {author} {\bibfnamefont {B.}~\bibnamefont {Liu}}, \bibinfo
  {author} {\bibfnamefont {W.}~\bibnamefont {Tian}}, \ and\ \bibinfo {author}
  {\bibfnamefont {T.}~\bibnamefont {Cui}},\ }\href {\doibase
  http://dx.doi.org/10.1038/srep06968} {\bibfield  {journal} {\bibinfo
  {journal} {Sci. Rep.}\ }\textbf {\bibinfo {volume} {4}} (\bibinfo {year}
  {2014}),\ http://dx.doi.org/10.1038/srep06968}\BibitemShut {NoStop}%
\bibitem [{\citenamefont {Drozdov}\ \emph
  {et~al.}(2015{\natexlab{b}})\citenamefont {Drozdov}, \citenamefont
  {Eremets},\ and\ \citenamefont {Troyan}}]{Drozdov_PH3_arxiv2015}%
  \BibitemOpen
  \bibfield  {author} {\bibinfo {author} {\bibfnamefont {A.}~\bibnamefont
  {Drozdov}}, \bibinfo {author} {\bibfnamefont {M.~I.}\ \bibnamefont
  {Eremets}}, \ and\ \bibinfo {author} {\bibfnamefont {I.~A.}\ \bibnamefont
  {Troyan}},\ }\href@noop {} {\bibfield  {journal} {\bibinfo  {journal} {ArXiv
  e-prints}\ } (\bibinfo {year} {2015}{\natexlab{b}})}\BibitemShut {NoStop}%
\bibitem [{\citenamefont {Shamp}\ \emph {et~al.}(2016)\citenamefont {Shamp},
  \citenamefont {Terpstra}, \citenamefont {Bi}, \citenamefont {Falls},
  \citenamefont {Avery},\ and\ \citenamefont
  {Zurek}}]{shamp_decomposition_2015}%
  \BibitemOpen
  \bibfield  {author} {\bibinfo {author} {\bibfnamefont {A.}~\bibnamefont
  {Shamp}}, \bibinfo {author} {\bibfnamefont {T.}~\bibnamefont {Terpstra}},
  \bibinfo {author} {\bibfnamefont {T.}~\bibnamefont {Bi}}, \bibinfo {author}
  {\bibfnamefont {Z.}~\bibnamefont {Falls}}, \bibinfo {author} {\bibfnamefont
  {P.}~\bibnamefont {Avery}}, \ and\ \bibinfo {author} {\bibfnamefont
  {E.}~\bibnamefont {Zurek}},\ }\href {\doibase 10.1021/jacs.5b10180}
  {\bibfield  {journal} {\bibinfo  {journal} {Journal of the American Chemical
  Society}\ }\textbf {\bibinfo {volume} {138}},\ \bibinfo {pages} {1884}
  (\bibinfo {year} {2016})},\ \bibinfo {note} {pMID: 26777416}\BibitemShut
  {NoStop}%
\bibitem [{\citenamefont {Fu}\ \emph {et~al.}(2016)\citenamefont {Fu},
  \citenamefont {Du}, \citenamefont {Zhang}, \citenamefont {Peng},
  \citenamefont {Zhang}, \citenamefont {Pickard}, \citenamefont {Needs},
  \citenamefont {Singh}, \citenamefont {Zheng},\ and\ \citenamefont
  {Ma}}]{Fu_Ma_pnictogenH_2016}%
  \BibitemOpen
  \bibfield  {author} {\bibinfo {author} {\bibfnamefont {Y.}~\bibnamefont
  {Fu}}, \bibinfo {author} {\bibfnamefont {X.}~\bibnamefont {Du}}, \bibinfo
  {author} {\bibfnamefont {L.}~\bibnamefont {Zhang}}, \bibinfo {author}
  {\bibfnamefont {F.}~\bibnamefont {Peng}}, \bibinfo {author} {\bibfnamefont
  {M.}~\bibnamefont {Zhang}}, \bibinfo {author} {\bibfnamefont {C.~J.}\
  \bibnamefont {Pickard}}, \bibinfo {author} {\bibfnamefont {R.~J.}\
  \bibnamefont {Needs}}, \bibinfo {author} {\bibfnamefont {D.~J.}\ \bibnamefont
  {Singh}}, \bibinfo {author} {\bibfnamefont {W.}~\bibnamefont {Zheng}}, \ and\
  \bibinfo {author} {\bibfnamefont {Y.}~\bibnamefont {Ma}},\ }\href {\doibase
  10.1021/acs.chemmater.5b04638} {\bibfield  {journal} {\bibinfo  {journal}
  {Chemistry of Materials}\ }\textbf {\bibinfo {volume} {28}},\ \bibinfo
  {pages} {1746} (\bibinfo {year} {2016})}\BibitemShut {NoStop}%
\bibitem [{\citenamefont {Flores-Livas}\ \emph
  {et~al.}(2016{\natexlab{a}})\citenamefont {Flores-Livas}, \citenamefont
  {Amsler}, \citenamefont {Heil}, \citenamefont {Sanna}, \citenamefont {Boeri},
  \citenamefont {Profeta}, \citenamefont {Wolverton}, \citenamefont
  {Goedecker},\ and\ \citenamefont {Gross}}]{Flores_PH3_PRBR2016}%
  \BibitemOpen
  \bibfield  {author} {\bibinfo {author} {\bibfnamefont {J.~A.}\ \bibnamefont
  {Flores-Livas}}, \bibinfo {author} {\bibfnamefont {M.}~\bibnamefont
  {Amsler}}, \bibinfo {author} {\bibfnamefont {C.}~\bibnamefont {Heil}},
  \bibinfo {author} {\bibfnamefont {A.}~\bibnamefont {Sanna}}, \bibinfo
  {author} {\bibfnamefont {L.}~\bibnamefont {Boeri}}, \bibinfo {author}
  {\bibfnamefont {G.}~\bibnamefont {Profeta}}, \bibinfo {author} {\bibfnamefont
  {C.}~\bibnamefont {Wolverton}}, \bibinfo {author} {\bibfnamefont
  {S.}~\bibnamefont {Goedecker}}, \ and\ \bibinfo {author} {\bibfnamefont
  {E.~K.~U.}\ \bibnamefont {Gross}},\ }\href {\doibase
  10.1103/PhysRevB.93.020508} {\bibfield  {journal} {\bibinfo  {journal} {Phys.
  Rev. B}\ }\textbf {\bibinfo {volume} {93}},\ \bibinfo {pages} {020508}
  (\bibinfo {year} {2016}{\natexlab{a}})}\BibitemShut {NoStop}%
\bibitem [{\citenamefont {Zurek}\ \emph {et~al.}(2009)\citenamefont {Zurek},
  \citenamefont {Hoffmann}, \citenamefont {Ashcroft}, \citenamefont {Oganov},\
  and\ \citenamefont {Lyakhov}}]{zurek2009little}%
  \BibitemOpen
  \bibfield  {author} {\bibinfo {author} {\bibfnamefont {E.}~\bibnamefont
  {Zurek}}, \bibinfo {author} {\bibfnamefont {R.}~\bibnamefont {Hoffmann}},
  \bibinfo {author} {\bibfnamefont {N.}~\bibnamefont {Ashcroft}}, \bibinfo
  {author} {\bibfnamefont {A.~R.}\ \bibnamefont {Oganov}}, \ and\ \bibinfo
  {author} {\bibfnamefont {A.~O.}\ \bibnamefont {Lyakhov}},\ }\href
  {http://www.pnas.org/content/106/42/17640.short} {\bibfield  {journal}
  {\bibinfo  {journal} {Proceedings of the National Academy of Sciences}\
  }\textbf {\bibinfo {volume} {106}},\ \bibinfo {pages} {17640} (\bibinfo
  {year} {2009})}\BibitemShut {NoStop}%
\bibitem [{\citenamefont {Wang}\ \emph {et~al.}(2012)\citenamefont {Wang},
  \citenamefont {Tse}, \citenamefont {Tanaka}, \citenamefont {Iitaka},\ and\
  \citenamefont {Ma}}]{Wang_PNAS2012_CaH6}%
  \BibitemOpen
  \bibfield  {author} {\bibinfo {author} {\bibfnamefont {H.}~\bibnamefont
  {Wang}}, \bibinfo {author} {\bibfnamefont {J.~S.}\ \bibnamefont {Tse}},
  \bibinfo {author} {\bibfnamefont {K.}~\bibnamefont {Tanaka}}, \bibinfo
  {author} {\bibfnamefont {T.}~\bibnamefont {Iitaka}}, \ and\ \bibinfo {author}
  {\bibfnamefont {Y.}~\bibnamefont {Ma}},\ }\href {\doibase
  10.1073/pnas.1118168109} {\bibfield  {journal} {\bibinfo  {journal}
  {Proceedings of the National Academy of Sciences}\ }\textbf {\bibinfo
  {volume} {109}},\ \bibinfo {pages} {6463} (\bibinfo {year}
  {2012})}\BibitemShut {NoStop}%
\bibitem [{\citenamefont {Li}\ \emph {et~al.}(2015)\citenamefont {Li},
  \citenamefont {Hao}, \citenamefont {Liu}, \citenamefont {Tse}, \citenamefont
  {Wang},\ and\ \citenamefont {Ma}}]{Li_SREP2015_YH6}%
  \BibitemOpen
  \bibfield  {author} {\bibinfo {author} {\bibfnamefont {Y.}~\bibnamefont
  {Li}}, \bibinfo {author} {\bibfnamefont {J.}~\bibnamefont {Hao}}, \bibinfo
  {author} {\bibfnamefont {H.}~\bibnamefont {Liu}}, \bibinfo {author}
  {\bibfnamefont {J.~S.}\ \bibnamefont {Tse}}, \bibinfo {author} {\bibfnamefont
  {Y.}~\bibnamefont {Wang}}, \ and\ \bibinfo {author} {\bibfnamefont
  {Y.}~\bibnamefont {Ma}},\ }\href {http://dx.doi.org/10.1038/srep09948}
  {\bibfield  {journal} {\bibinfo  {journal} {Scientific Reports}\ }\textbf
  {\bibinfo {volume} {5}},\ \bibinfo {pages} {9948 EP } (\bibinfo {year}
  {2015})}\BibitemShut {NoStop}%
\bibitem [{\citenamefont {Struzhkin}(2015)}]{Struzhkin_rev}%
  \BibitemOpen
  \bibfield  {author} {\bibinfo {author} {\bibfnamefont {V.~V.}\ \bibnamefont
  {Struzhkin}},\ }\href {\doibase
  http://dx.doi.org/10.1016/j.physc.2015.02.017} {\bibfield  {journal}
  {\bibinfo  {journal} {Physica C}\ }\textbf {\bibinfo {volume} {514}},\
  \bibinfo {pages} {77 } (\bibinfo {year} {2015})}\BibitemShut {NoStop}%
\bibitem [{\citenamefont {Flores-Livas}\ \emph
  {et~al.}(2016{\natexlab{b}})\citenamefont {Flores-Livas}, \citenamefont
  {Sanna}, \citenamefont {Davydov}, \citenamefont {Goedecker},\ and\
  \citenamefont {Marques}}]{Flores_H2O}%
  \BibitemOpen
  \bibfield  {author} {\bibinfo {author} {\bibfnamefont {J.~A.}\ \bibnamefont
  {Flores-Livas}}, \bibinfo {author} {\bibfnamefont {A.}~\bibnamefont {Sanna}},
  \bibinfo {author} {\bibfnamefont {A.}~\bibnamefont {Davydov}}, \bibinfo
  {author} {\bibfnamefont {S.}~\bibnamefont {Goedecker}}, \ and\ \bibinfo
  {author} {\bibfnamefont {M.~A.~L.}\ \bibnamefont {Marques}},\ }\href
  {arXiv:1610.04110} {\bibfield  {journal} {\bibinfo  {journal}
  {airxiv-condmat/1610.04110}\ } (\bibinfo {year}
  {2016}{\natexlab{b}})}\BibitemShut {NoStop}%
\bibitem [{\citenamefont {Shamp}\ and\ \citenamefont
  {Zurek}(2015)}]{Zurek_IodineH_JPCL2016}%
  \BibitemOpen
  \bibfield  {author} {\bibinfo {author} {\bibfnamefont {A.}~\bibnamefont
  {Shamp}}\ and\ \bibinfo {author} {\bibfnamefont {E.}~\bibnamefont {Zurek}},\
  }\href {\doibase 10.1021/acs.jpclett.5b01839} {\bibfield  {journal} {\bibinfo
   {journal} {The Journal of Physical Chemistry Letters}\ }\textbf {\bibinfo
  {volume} {6}},\ \bibinfo {pages} {4067} (\bibinfo {year} {2015})},\ \bibinfo
  {note} {pMID: 26722778}\BibitemShut {NoStop}%
\bibitem [{\citenamefont {Kim}\ \emph {et~al.}(2010)\citenamefont {Kim},
  \citenamefont {Scheicher}, \citenamefont {Mao}, \citenamefont {Kang},\ and\
  \citenamefont {Ahuja}}]{kim_general_2010}%
  \BibitemOpen
  \bibfield  {author} {\bibinfo {author} {\bibfnamefont {D.~Y.}\ \bibnamefont
  {Kim}}, \bibinfo {author} {\bibfnamefont {R.~H.}\ \bibnamefont {Scheicher}},
  \bibinfo {author} {\bibfnamefont {H.-k.}\ \bibnamefont {Mao}}, \bibinfo
  {author} {\bibfnamefont {T.~W.}\ \bibnamefont {Kang}}, \ and\ \bibinfo
  {author} {\bibfnamefont {R.}~\bibnamefont {Ahuja}},\ }\href {\doibase
  10.1073/pnas.0914462107} {\bibfield  {journal} {\bibinfo  {journal} {PNAS}\
  }\textbf {\bibinfo {volume} {107}},\ \bibinfo {pages} {2793} (\bibinfo {year}
  {2010})}\BibitemShut {NoStop}%
\bibitem [{\citenamefont {Bernstein}\ \emph {et~al.}(2015)\citenamefont
  {Bernstein}, \citenamefont {Hellberg}, \citenamefont {Johannes},
  \citenamefont {Mazin},\ and\ \citenamefont {Mehl}}]{SH_PRB-Mazin-2015}%
  \BibitemOpen
  \bibfield  {author} {\bibinfo {author} {\bibfnamefont {N.}~\bibnamefont
  {Bernstein}}, \bibinfo {author} {\bibfnamefont {C.~S.}\ \bibnamefont
  {Hellberg}}, \bibinfo {author} {\bibfnamefont {M.~D.}\ \bibnamefont
  {Johannes}}, \bibinfo {author} {\bibfnamefont {I.~I.}\ \bibnamefont {Mazin}},
  \ and\ \bibinfo {author} {\bibfnamefont {M.~J.}\ \bibnamefont {Mehl}},\
  }\href {\doibase 10.1103/PhysRevB.91.060511} {\bibfield  {journal} {\bibinfo
  {journal} {Phys. Rev. B}\ }\textbf {\bibinfo {volume} {91}},\ \bibinfo
  {pages} {060511} (\bibinfo {year} {2015})}\BibitemShut {NoStop}%
\bibitem [{\citenamefont {Heil}\ and\ \citenamefont
  {Boeri}(2015)}]{Heil-Boeri_PRB2015}%
  \BibitemOpen
  \bibfield  {author} {\bibinfo {author} {\bibfnamefont {C.}~\bibnamefont
  {Heil}}\ and\ \bibinfo {author} {\bibfnamefont {L.}~\bibnamefont {Boeri}},\
  }\href {\doibase 10.1103/PhysRevB.92.060508} {\bibfield  {journal} {\bibinfo
  {journal} {Phys. Rev. B}\ }\textbf {\bibinfo {volume} {92}},\ \bibinfo
  {pages} {060508} (\bibinfo {year} {2015})}\BibitemShut {NoStop}%
\bibitem [{\citenamefont {Flores-Livas}\ \emph
  {et~al.}(2016{\natexlab{c}})\citenamefont {Flores-Livas}, \citenamefont
  {Sanna},\ and\ \citenamefont {Gross}}]{Flores-Livas2016}%
  \BibitemOpen
  \bibfield  {author} {\bibinfo {author} {\bibfnamefont {J.~A.}\ \bibnamefont
  {Flores-Livas}}, \bibinfo {author} {\bibfnamefont {A.}~\bibnamefont {Sanna}},
  \ and\ \bibinfo {author} {\bibfnamefont {E.~K.}\ \bibnamefont {Gross}},\
  }\href {\doibase 10.1140/epjb/e2016-70020-0} {\bibfield  {journal} {\bibinfo
  {journal} {The European Physical Journal B}\ }\textbf {\bibinfo {volume}
  {89}},\ \bibinfo {pages} {63} (\bibinfo {year}
  {2016}{\natexlab{c}})}\BibitemShut {NoStop}%
\bibitem [{\citenamefont {Quan}\ and\ \citenamefont
  {Pickett}(2016)}]{PhysRevB.93.104526}%
  \BibitemOpen
  \bibfield  {author} {\bibinfo {author} {\bibfnamefont {Y.}~\bibnamefont
  {Quan}}\ and\ \bibinfo {author} {\bibfnamefont {W.~E.}\ \bibnamefont
  {Pickett}},\ }\href {\doibase 10.1103/PhysRevB.93.104526} {\bibfield
  {journal} {\bibinfo  {journal} {Phys. Rev. B}\ }\textbf {\bibinfo {volume}
  {93}},\ \bibinfo {pages} {104526} (\bibinfo {year} {2016})}\BibitemShut
  {NoStop}%
\bibitem [{\citenamefont {Li}\ \emph {et~al.}(2014)\citenamefont {Li},
  \citenamefont {Hao}, \citenamefont {Liu}, \citenamefont {Li},\ and\
  \citenamefont {Ma}}]{doi:10.1063/1.4874158}%
  \BibitemOpen
  \bibfield  {author} {\bibinfo {author} {\bibfnamefont {Y.}~\bibnamefont
  {Li}}, \bibinfo {author} {\bibfnamefont {J.}~\bibnamefont {Hao}}, \bibinfo
  {author} {\bibfnamefont {H.}~\bibnamefont {Liu}}, \bibinfo {author}
  {\bibfnamefont {Y.}~\bibnamefont {Li}}, \ and\ \bibinfo {author}
  {\bibfnamefont {Y.}~\bibnamefont {Ma}},\ }\href {\doibase 10.1063/1.4874158}
  {\bibfield  {journal} {\bibinfo  {journal} {The Journal of Chemical Physics}\
  }\textbf {\bibinfo {volume} {140}},\ \bibinfo {pages} {174712} (\bibinfo
  {year} {2014})}\BibitemShut {NoStop}%
\bibitem [{\citenamefont {Papaconstantopoulos}\ \emph
  {et~al.}(2015)\citenamefont {Papaconstantopoulos}, \citenamefont {Klein},
  \citenamefont {Mehl},\ and\ \citenamefont {Pickett}}]{PhysRevB.91.184511}%
  \BibitemOpen
  \bibfield  {author} {\bibinfo {author} {\bibfnamefont {D.~A.}\ \bibnamefont
  {Papaconstantopoulos}}, \bibinfo {author} {\bibfnamefont {B.~M.}\
  \bibnamefont {Klein}}, \bibinfo {author} {\bibfnamefont {M.~J.}\ \bibnamefont
  {Mehl}}, \ and\ \bibinfo {author} {\bibfnamefont {W.~E.}\ \bibnamefont
  {Pickett}},\ }\href {\doibase 10.1103/PhysRevB.91.184511} {\bibfield
  {journal} {\bibinfo  {journal} {Phys. Rev. B}\ }\textbf {\bibinfo {volume}
  {91}},\ \bibinfo {pages} {184511} (\bibinfo {year} {2015})}\BibitemShut
  {NoStop}%
\bibitem [{\citenamefont {Ortenzi}\ \emph {et~al.}(2016)\citenamefont
  {Ortenzi}, \citenamefont {Cappelluti},\ and\ \citenamefont
  {Pietronero}}]{PhysRevB.94.064507}%
  \BibitemOpen
  \bibfield  {author} {\bibinfo {author} {\bibfnamefont {L.}~\bibnamefont
  {Ortenzi}}, \bibinfo {author} {\bibfnamefont {E.}~\bibnamefont {Cappelluti}},
  \ and\ \bibinfo {author} {\bibfnamefont {L.}~\bibnamefont {Pietronero}},\
  }\href {\doibase 10.1103/PhysRevB.94.064507} {\bibfield  {journal} {\bibinfo
  {journal} {Phys. Rev. B}\ }\textbf {\bibinfo {volume} {94}},\ \bibinfo
  {pages} {064507} (\bibinfo {year} {2016})}\BibitemShut {NoStop}%
\bibitem [{\citenamefont {Errea}\ \emph {et~al.}(2015)\citenamefont {Errea},
  \citenamefont {Calandra}, \citenamefont {Pickard}, \citenamefont {Nelson},
  \citenamefont {Needs}, \citenamefont {Li}, \citenamefont {Liu}, \citenamefont
  {Zhang}, \citenamefont {Ma},\ and\ \citenamefont
  {Mauri}}]{PhysRevLett.114.157004}%
  \BibitemOpen
  \bibfield  {author} {\bibinfo {author} {\bibfnamefont {I.}~\bibnamefont
  {Errea}}, \bibinfo {author} {\bibfnamefont {M.}~\bibnamefont {Calandra}},
  \bibinfo {author} {\bibfnamefont {C.~J.}\ \bibnamefont {Pickard}}, \bibinfo
  {author} {\bibfnamefont {J.}~\bibnamefont {Nelson}}, \bibinfo {author}
  {\bibfnamefont {R.~J.}\ \bibnamefont {Needs}}, \bibinfo {author}
  {\bibfnamefont {Y.}~\bibnamefont {Li}}, \bibinfo {author} {\bibfnamefont
  {H.}~\bibnamefont {Liu}}, \bibinfo {author} {\bibfnamefont {Y.}~\bibnamefont
  {Zhang}}, \bibinfo {author} {\bibfnamefont {Y.}~\bibnamefont {Ma}}, \ and\
  \bibinfo {author} {\bibfnamefont {F.}~\bibnamefont {Mauri}},\ }\href
  {\doibase 10.1103/PhysRevLett.114.157004} {\bibfield  {journal} {\bibinfo
  {journal} {Phys. Rev. Lett.}\ }\textbf {\bibinfo {volume} {114}},\ \bibinfo
  {pages} {157004} (\bibinfo {year} {2015})}\BibitemShut {NoStop}%
\bibitem [{\citenamefont {{Schlapbach Louis}}\ and\ \citenamefont {{Zuttel
  Andreas}}(2001)}]{nature_review_hydrogen}%
  \BibitemOpen
  \bibfield  {author} {\bibinfo {author} {\bibnamefont {{Schlapbach Louis}}}\
  and\ \bibinfo {author} {\bibnamefont {{Zuttel Andreas}}},\ }\href {\doibase
  http://dx.doi.org/10.1038/35104634} {\bibfield  {journal} {\bibinfo
  {journal} {Nature}\ }\textbf {\bibinfo {volume} {414}},\ \bibinfo {pages}
  {353} (\bibinfo {year} {2001})},\ \bibinfo {note}
  {10.1038/35104634}\BibitemShut {NoStop}%
\bibitem [{\citenamefont {Akbarzadeh}\ \emph {et~al.}(2007)\citenamefont
  {Akbarzadeh}, \citenamefont {V.},\ and\ \citenamefont
  {C.}}]{wolverton_advmat_2007}%
  \BibitemOpen
  \bibfield  {author} {\bibinfo {author} {\bibfnamefont {R.}~\bibnamefont
  {Akbarzadeh}}, \bibinfo {author} {\bibfnamefont {O.}~\bibnamefont {V.}}, \
  and\ \bibinfo {author} {\bibfnamefont {W.}~\bibnamefont {C.}},\ }\href
  {\doibase 10.1002/adma.200700843} {\bibfield  {journal} {\bibinfo  {journal}
  {Advanced Materials}\ }\textbf {\bibinfo {volume} {19}},\ \bibinfo {pages}
  {3233} (\bibinfo {year} {2007})}\BibitemShut {NoStop}%
\bibitem [{\citenamefont {Baroni}\ \emph {et~al.}(2001)\citenamefont {Baroni},
  \citenamefont {de~Gironcoli}, \citenamefont {Dal~Corso},\ and\ \citenamefont
  {Giannozzi}}]{Baroni_RMP2001}%
  \BibitemOpen
  \bibfield  {author} {\bibinfo {author} {\bibfnamefont {S.}~\bibnamefont
  {Baroni}}, \bibinfo {author} {\bibfnamefont {S.}~\bibnamefont
  {de~Gironcoli}}, \bibinfo {author} {\bibfnamefont {A.}~\bibnamefont
  {Dal~Corso}}, \ and\ \bibinfo {author} {\bibfnamefont {P.}~\bibnamefont
  {Giannozzi}},\ }\href {\doibase 10.1103/RevModPhys.73.515} {\bibfield
  {journal} {\bibinfo  {journal} {Rev. Mod. Phys.}\ }\textbf {\bibinfo {volume}
  {73}},\ \bibinfo {pages} {515} (\bibinfo {year} {2001})}\BibitemShut
  {NoStop}%
\bibitem [{\citenamefont {Giannozzi}\ \emph {et~al.}(2009)\citenamefont
  {Giannozzi}, \citenamefont {Baroni}, \citenamefont {Bonini}, \citenamefont
  {Calandra}, \citenamefont {Car}, \citenamefont {Cavazzoni}, \citenamefont
  {Ceresoli}, \citenamefont {Chiarotti}, \citenamefont {Cococcioni},
  \citenamefont {Dabo}, \citenamefont {Corso}, \citenamefont {de~Gironcoli},
  \citenamefont {Fabris}, \citenamefont {Fratesi}, \citenamefont {Gebauer},
  \citenamefont {Gerstmann}, \citenamefont {Gougoussis}, \citenamefont
  {Kokalj}, \citenamefont {Lazzeri}, \citenamefont {Martin-Samos},
  \citenamefont {Marzari}, \citenamefont {Mauri}, \citenamefont {Mazzarello},
  \citenamefont {Paolini}, \citenamefont {Pasquarello}, \citenamefont
  {Paulatto}, \citenamefont {Sbraccia}, \citenamefont {Scandolo}, \citenamefont
  {Sclauzero}, \citenamefont {Seitsonen}, \citenamefont {Smogunov},
  \citenamefont {Umari},\ and\ \citenamefont {Wentzcovitch}}]{QE_JPCM_2009}%
  \BibitemOpen
  \bibfield  {author} {\bibinfo {author} {\bibfnamefont {P.}~\bibnamefont
  {Giannozzi}}, \bibinfo {author} {\bibfnamefont {S.}~\bibnamefont {Baroni}},
  \bibinfo {author} {\bibfnamefont {N.}~\bibnamefont {Bonini}}, \bibinfo
  {author} {\bibfnamefont {M.}~\bibnamefont {Calandra}}, \bibinfo {author}
  {\bibfnamefont {R.}~\bibnamefont {Car}}, \bibinfo {author} {\bibfnamefont
  {C.}~\bibnamefont {Cavazzoni}}, \bibinfo {author} {\bibfnamefont
  {D.}~\bibnamefont {Ceresoli}}, \bibinfo {author} {\bibfnamefont {G.~L.}\
  \bibnamefont {Chiarotti}}, \bibinfo {author} {\bibfnamefont {M.}~\bibnamefont
  {Cococcioni}}, \bibinfo {author} {\bibfnamefont {I.}~\bibnamefont {Dabo}},
  \bibinfo {author} {\bibfnamefont {A.~D.}\ \bibnamefont {Corso}}, \bibinfo
  {author} {\bibfnamefont {S.}~\bibnamefont {de~Gironcoli}}, \bibinfo {author}
  {\bibfnamefont {S.}~\bibnamefont {Fabris}}, \bibinfo {author} {\bibfnamefont
  {G.}~\bibnamefont {Fratesi}}, \bibinfo {author} {\bibfnamefont
  {R.}~\bibnamefont {Gebauer}}, \bibinfo {author} {\bibfnamefont
  {U.}~\bibnamefont {Gerstmann}}, \bibinfo {author} {\bibfnamefont
  {C.}~\bibnamefont {Gougoussis}}, \bibinfo {author} {\bibfnamefont
  {A.}~\bibnamefont {Kokalj}}, \bibinfo {author} {\bibfnamefont
  {M.}~\bibnamefont {Lazzeri}}, \bibinfo {author} {\bibfnamefont
  {L.}~\bibnamefont {Martin-Samos}}, \bibinfo {author} {\bibfnamefont
  {N.}~\bibnamefont {Marzari}}, \bibinfo {author} {\bibfnamefont
  {F.}~\bibnamefont {Mauri}}, \bibinfo {author} {\bibfnamefont
  {R.}~\bibnamefont {Mazzarello}}, \bibinfo {author} {\bibfnamefont
  {S.}~\bibnamefont {Paolini}}, \bibinfo {author} {\bibfnamefont
  {A.}~\bibnamefont {Pasquarello}}, \bibinfo {author} {\bibfnamefont
  {L.}~\bibnamefont {Paulatto}}, \bibinfo {author} {\bibfnamefont
  {C.}~\bibnamefont {Sbraccia}}, \bibinfo {author} {\bibfnamefont
  {S.}~\bibnamefont {Scandolo}}, \bibinfo {author} {\bibfnamefont
  {G.}~\bibnamefont {Sclauzero}}, \bibinfo {author} {\bibfnamefont {A.~P.}\
  \bibnamefont {Seitsonen}}, \bibinfo {author} {\bibfnamefont {A.}~\bibnamefont
  {Smogunov}}, \bibinfo {author} {\bibfnamefont {P.}~\bibnamefont {Umari}}, \
  and\ \bibinfo {author} {\bibfnamefont {R.~M.}\ \bibnamefont {Wentzcovitch}},\
  }\href {http://stacks.iop.org/0953-8984/21/i=39/a=395502} {\bibfield
  {journal} {\bibinfo  {journal} {Journal of Physics: Condensed Matter}\
  }\textbf {\bibinfo {volume} {21}},\ \bibinfo {pages} {395502} (\bibinfo
  {year} {2009})}\BibitemShut {NoStop}%
\bibitem [{\citenamefont {Z{\"u}ttel}\ \emph {et~al.}(2003)\citenamefont
  {Z{\"u}ttel}, \citenamefont {Wenger}, \citenamefont {Rentsch}, \citenamefont
  {Sudan}, , \citenamefont {Mauron},\ and\ \citenamefont
  {Emmenegger}}]{scidirzuet2003}%
  \BibitemOpen
  \bibfield  {author} {\bibinfo {author} {\bibfnamefont {A.}~\bibnamefont
  {Z{\"u}ttel}}, \bibinfo {author} {\bibfnamefont {P.}~\bibnamefont {Wenger}},
  \bibinfo {author} {\bibfnamefont {S.}~\bibnamefont {Rentsch}}, \bibinfo
  {author} {\bibfnamefont {P.}~\bibnamefont {Sudan}}, , \bibinfo {author}
  {\bibfnamefont {P.}~\bibnamefont {Mauron}}, \ and\ \bibinfo {author}
  {\bibfnamefont {C.}~\bibnamefont {Emmenegger}},\ }\href
  {http://www.sciencedirect.com/science/article/pii/S0378775303000545}
  {\bibfield  {journal} {\bibinfo  {journal} {Journal of Power Sources}\
  }\textbf {\bibinfo {volume} {118}},\ \bibinfo {pages} {1–7} (\bibinfo
  {year} {2003})}\BibitemShut {NoStop}%
\bibitem [{\citenamefont {{Vajo John J.}}\ \emph {et~al.}(2005)\citenamefont
  {{Vajo John J.}}, \citenamefont {{Skeith Sky L.}},\ and\ \citenamefont
  {{Mertens Florian}}}]{vajo2005}%
  \BibitemOpen
  \bibfield  {author} {\bibinfo {author} {\bibnamefont {{Vajo John J.}}},
  \bibinfo {author} {\bibnamefont {{Skeith Sky L.}}}, \ and\ \bibinfo {author}
  {\bibnamefont {{Mertens Florian}}},\ }\href {\doibase
  http://dx.doi.org/10.1021/jp040769o} {\bibfield  {journal} {\bibinfo
  {journal} {The Journal of Physical Chemistry B}\ }\textbf {\bibinfo {volume}
  {109}},\ \bibinfo {pages} {3719–3722} (\bibinfo {year} {2005})},\ \bibinfo
  {note} {doi: 10.1021/jp040769o}\BibitemShut {NoStop}%
\bibitem [{\citenamefont {{Z\"uttel A.}}\ \emph {et~al.}(2003)\citenamefont
  {{Z\"uttel A.}}, \citenamefont {{Rentsch S.}}, \citenamefont {{Fischer P.}},
  \citenamefont {{Wenger P.}}, \citenamefont {{Sudan P.}}, \citenamefont
  {{Mauron Ph.}},\ and\ \citenamefont {{Emmenegger Ch.}}}]{zuettel2003}%
  \BibitemOpen
  \bibfield  {author} {\bibinfo {author} {\bibnamefont {{Z\"uttel A.}}},
  \bibinfo {author} {\bibnamefont {{Rentsch S.}}}, \bibinfo {author}
  {\bibnamefont {{Fischer P.}}}, \bibinfo {author} {\bibnamefont {{Wenger
  P.}}}, \bibinfo {author} {\bibnamefont {{Sudan P.}}}, \bibinfo {author}
  {\bibnamefont {{Mauron Ph.}}}, \ and\ \bibinfo {author} {\bibnamefont
  {{Emmenegger Ch.}}},\ }\href
  {http://www.sciencedirect.com/science/article/pii/S0925838802012537}
  {\bibfield  {journal} {\bibinfo  {journal} {Journal of Alloys and Compounds}\
  }\textbf {\bibinfo {volume} {356-357}},\ \bibinfo {pages} {515} (\bibinfo
  {year} {2003})}\BibitemShut {NoStop}%
\bibitem [{\citenamefont {Miwa}\ \emph {et~al.}(2004)\citenamefont {Miwa},
  \citenamefont {Ohba}, \citenamefont {Towata}, \citenamefont {Nakamori},\ and\
  \citenamefont {Orimo}}]{PhysRevB.69.245120}%
  \BibitemOpen
  \bibfield  {author} {\bibinfo {author} {\bibfnamefont {K.}~\bibnamefont
  {Miwa}}, \bibinfo {author} {\bibfnamefont {N.}~\bibnamefont {Ohba}}, \bibinfo
  {author} {\bibfnamefont {S.-i.}\ \bibnamefont {Towata}}, \bibinfo {author}
  {\bibfnamefont {Y.}~\bibnamefont {Nakamori}}, \ and\ \bibinfo {author}
  {\bibfnamefont {S.-i.}\ \bibnamefont {Orimo}},\ }\href {\doibase
  10.1103/PhysRevB.69.245120} {\bibfield  {journal} {\bibinfo  {journal} {Phys.
  Rev. B}\ }\textbf {\bibinfo {volume} {69}},\ \bibinfo {pages} {245120}
  (\bibinfo {year} {2004})}\BibitemShut {NoStop}%
\bibitem [{\citenamefont {Orimo}\ and\ \citenamefont
  {Nakamori}(2006)}]{shinorimo2006}%
  \BibitemOpen
  \bibfield  {author} {\bibinfo {author} {\bibfnamefont {S.-I.}\ \bibnamefont
  {Orimo}}\ and\ \bibinfo {author} {\bibfnamefont {Y.}~\bibnamefont
  {Nakamori}},\ }\href {\doibase http://dx.doi.org/10.1063/1.2221880}
  {\bibfield  {journal} {\bibinfo  {journal} {Applied Physics Letters}\
  }\textbf {\bibinfo {volume} {89}},\ \bibinfo {pages} {021920} (\bibinfo
  {year} {2006})},\ \bibinfo {note} {doi: 10.1063/1.2221880}\BibitemShut
  {NoStop}%
\bibitem [{\citenamefont {Souli\'e}\ \emph {et~al.}(2002)\citenamefont
  {Souli\'e}, \citenamefont {Renaudin}, \citenamefont {Cern\'y},\ and\
  \citenamefont {Yvon}}]{soulie2002}%
  \BibitemOpen
  \bibfield  {author} {\bibinfo {author} {\bibfnamefont {J.-P.}\ \bibnamefont
  {Souli\'e}}, \bibinfo {author} {\bibfnamefont {G.}~\bibnamefont {Renaudin}},
  \bibinfo {author} {\bibfnamefont {R.}~\bibnamefont {Cern\'y}}, \ and\
  \bibinfo {author} {\bibfnamefont {K.}~\bibnamefont {Yvon}},\ }\href
  {http://www.sciencedirect.com/science/article/pii/S0925838802005212}
  {\bibfield  {journal} {\bibinfo  {journal} {Journal of Alloys and Compounds}\
  }\textbf {\bibinfo {volume} {346}},\ \bibinfo {pages} {200} (\bibinfo {year}
  {2002})}\BibitemShut {NoStop}%
\bibitem [{\citenamefont {{Gross Adam F.}}\ \emph {et~al.}(2008)\citenamefont
  {{Gross Adam F.}}, \citenamefont {{Vajo John J.}}, \citenamefont {{Van Atta
  Sky L.}},\ and\ \citenamefont {{Olson Gregory L.}}}]{gross2008}%
  \BibitemOpen
  \bibfield  {author} {\bibinfo {author} {\bibnamefont {{Gross Adam F.}}},
  \bibinfo {author} {\bibnamefont {{Vajo John J.}}}, \bibinfo {author}
  {\bibnamefont {{Van Atta Sky L.}}}, \ and\ \bibinfo {author} {\bibnamefont
  {{Olson Gregory L.}}},\ }\href {\doibase http://dx.doi.org/10.1021/jp711066t}
  {\bibfield  {journal} {\bibinfo  {journal} {The Journal of Physical Chemistry
  C}\ }\textbf {\bibinfo {volume} {112}},\ \bibinfo {pages} {5651} (\bibinfo
  {year} {2008})},\ \bibinfo {note} {doi: 10.1021/jp711066t}\BibitemShut
  {NoStop}%
\bibitem [{\citenamefont {Mauron}\ \emph {et~al.}(2008)\citenamefont {Mauron},
  \citenamefont {Florian}, \citenamefont {Friedrichs}, \citenamefont {Remhof},
  \citenamefont {Bielmann}, \citenamefont {aZwicky},\ and\ \citenamefont
  {Z{\"u}ttel}}]{mauronacs2008}%
  \BibitemOpen
  \bibfield  {author} {\bibinfo {author} {\bibfnamefont {P.}~\bibnamefont
  {Mauron}}, \bibinfo {author} {\bibfnamefont {F.~B.}\ \bibnamefont {Florian}},
  \bibinfo {author} {\bibfnamefont {O.}~\bibnamefont {Friedrichs}}, \bibinfo
  {author} {\bibfnamefont {A.}~\bibnamefont {Remhof}}, \bibinfo {author}
  {\bibfnamefont {M.}~\bibnamefont {Bielmann}}, \bibinfo {author}
  {\bibfnamefont {C.~N.}\ \bibnamefont {aZwicky}}, \ and\ \bibinfo {author}
  {\bibfnamefont {A.}~\bibnamefont {Z{\"u}ttel}},\ }\href {\doibase
  http://dx.doi.org/10.1021/jp077572r} {\bibfield  {journal} {\bibinfo
  {journal} {The Journal of Physical Chemistry B}\ }\textbf {\bibinfo {volume}
  {112}},\ \bibinfo {pages} {906–910} (\bibinfo {year} {2008})},\ \bibinfo
  {note} {doi: 10.1021/jp077572r}\BibitemShut {NoStop}%
\bibitem [{\citenamefont {{Orimo S.}}\ \emph {et~al.}(2005)\citenamefont
  {{Orimo S.}}, \citenamefont {{Nakamori Y.}}, \citenamefont {{Kitahara G.}},
  \citenamefont {{Miwa K.}}, \citenamefont {{Ohba N.}}, \citenamefont {{Towata
  S.}},\ and\ \citenamefont {{Züttel A.}}}]{orimo2005}%
  \BibitemOpen
  \bibfield  {author} {\bibinfo {author} {\bibnamefont {{Orimo S.}}}, \bibinfo
  {author} {\bibnamefont {{Nakamori Y.}}}, \bibinfo {author} {\bibnamefont
  {{Kitahara G.}}}, \bibinfo {author} {\bibnamefont {{Miwa K.}}}, \bibinfo
  {author} {\bibnamefont {{Ohba N.}}}, \bibinfo {author} {\bibnamefont {{Towata
  S.}}}, \ and\ \bibinfo {author} {\bibnamefont {{Züttel A.}}},\ }\href
  {http://www.sciencedirect.com/science/article/pii/S0925838805009230}
  {\bibfield  {journal} {\bibinfo  {journal} {Journal of Alloys and Compounds}\
  }\textbf {\bibinfo {volume} {404-406}},\ \bibinfo {pages} {427} (\bibinfo
  {year} {2005})}\BibitemShut {NoStop}%
\bibitem [{\citenamefont {Kolmogorov}\ \emph {et~al.}(2015)\citenamefont
  {Kolmogorov}, \citenamefont {Hajinazar}, \citenamefont {Angyal},
  \citenamefont {Kuznetsov},\ and\ \citenamefont
  {Jephcoat}}]{kolmogorov_libh_2015}%
  \BibitemOpen
  \bibfield  {author} {\bibinfo {author} {\bibfnamefont {A.~N.}\ \bibnamefont
  {Kolmogorov}}, \bibinfo {author} {\bibfnamefont {S.}~\bibnamefont
  {Hajinazar}}, \bibinfo {author} {\bibfnamefont {C.}~\bibnamefont {Angyal}},
  \bibinfo {author} {\bibfnamefont {V.~L.}\ \bibnamefont {Kuznetsov}}, \ and\
  \bibinfo {author} {\bibfnamefont {A.~P.}\ \bibnamefont {Jephcoat}},\ }\href
  {\doibase 10.1103/PhysRevB.92.144110} {\bibfield  {journal} {\bibinfo
  {journal} {Phys. Rev. B}\ }\textbf {\bibinfo {volume} {92}},\ \bibinfo
  {pages} {144110} (\bibinfo {year} {2015})}\BibitemShut {NoStop}%
\bibitem [{\citenamefont {Juha}\ \emph {et~al.}(2007)\citenamefont {Juha},
  \citenamefont {Kirsi}, \citenamefont {Johanna}, \citenamefont {Elias},
  \citenamefont {Anssi},\ and\ \citenamefont {Alexander}}]{tuorinemi2007}%
  \BibitemOpen
  \bibfield  {author} {\bibinfo {author} {\bibfnamefont {T.}~\bibnamefont
  {Juha}}, \bibinfo {author} {\bibfnamefont {J.-N.}\ \bibnamefont {Kirsi}},
  \bibinfo {author} {\bibfnamefont {U.}~\bibnamefont {Johanna}}, \bibinfo
  {author} {\bibfnamefont {P.}~\bibnamefont {Elias}}, \bibinfo {author}
  {\bibfnamefont {S.}~\bibnamefont {Anssi}}, \ and\ \bibinfo {author}
  {\bibfnamefont {S.}~\bibnamefont {Alexander}},\ }\href {\doibase
  http://dx.doi.org/10.1038/nature05820} {\bibfield  {journal} {\bibinfo
  {journal} {Nature}\ }\textbf {\bibinfo {volume} {447}},\ \bibinfo {pages}
  {187} (\bibinfo {year} {2007})},\ \bibinfo {note}
  {10.1038/nature05820}\BibitemShut {NoStop}%
\bibitem [{\citenamefont {{Shimizu Katsuya}}\ \emph {et~al.}(2002)\citenamefont
  {{Shimizu Katsuya}}, \citenamefont {{Ishikawa Hiroto}}, \citenamefont {{Takao
  Daigoroh}}, \citenamefont {{Yagi Takehiko}},\ and\ \citenamefont {{Amaya
  Kiichi}}}]{shimizuli2002}%
  \BibitemOpen
  \bibfield  {author} {\bibinfo {author} {\bibnamefont {{Shimizu Katsuya}}},
  \bibinfo {author} {\bibnamefont {{Ishikawa Hiroto}}}, \bibinfo {author}
  {\bibnamefont {{Takao Daigoroh}}}, \bibinfo {author} {\bibnamefont {{Yagi
  Takehiko}}}, \ and\ \bibinfo {author} {\bibnamefont {{Amaya Kiichi}}},\
  }\href {\doibase http://dx.doi.org/10.1038/nature01098} {\bibfield  {journal}
  {\bibinfo  {journal} {Nature}\ }\textbf {\bibinfo {volume} {419}},\ \bibinfo
  {pages} {597–599} (\bibinfo {year} {2002})},\ \bibinfo {note}
  {10.1038/nature01098}\BibitemShut {NoStop}%
\bibitem [{\citenamefont {{Struzhkin Viktor V.}}\ \emph
  {et~al.}(2002)\citenamefont {{Struzhkin Viktor V.}}, \citenamefont {{Eremets
  Mikhail I.}}, \citenamefont {{Gan Wei}}, \citenamefont {{Mao Ho-kwang}},\
  and\ \citenamefont {{Hemley Russell J.}}}]{eremtsli2002}%
  \BibitemOpen
  \bibfield  {author} {\bibinfo {author} {\bibnamefont {{Struzhkin Viktor
  V.}}}, \bibinfo {author} {\bibnamefont {{Eremets Mikhail I.}}}, \bibinfo
  {author} {\bibnamefont {{Gan Wei}}}, \bibinfo {author} {\bibnamefont {{Mao
  Ho-kwang}}}, \ and\ \bibinfo {author} {\bibnamefont {{Hemley Russell J.}}},\
  }\href {http://science.sciencemag.org/content/298/5596/1213.abstract}
  {\bibfield  {journal} {\bibinfo  {journal} {Science}\ }\textbf {\bibinfo
  {volume} {298}},\ \bibinfo {pages} {1213} (\bibinfo {year}
  {2002})}\BibitemShut {NoStop}%
\bibitem [{\citenamefont {Lv}\ \emph {et~al.}(2011)\citenamefont {Lv},
  \citenamefont {Wang}, \citenamefont {Zhu},\ and\ \citenamefont
  {Ma}}]{PhysRevLett.106.015503}%
  \BibitemOpen
  \bibfield  {author} {\bibinfo {author} {\bibfnamefont {J.}~\bibnamefont
  {Lv}}, \bibinfo {author} {\bibfnamefont {Y.}~\bibnamefont {Wang}}, \bibinfo
  {author} {\bibfnamefont {L.}~\bibnamefont {Zhu}}, \ and\ \bibinfo {author}
  {\bibfnamefont {Y.}~\bibnamefont {Ma}},\ }\href {\doibase
  10.1103/PhysRevLett.106.015503} {\bibfield  {journal} {\bibinfo  {journal}
  {Phys. Rev. Lett.}\ }\textbf {\bibinfo {volume} {106}},\ \bibinfo {pages}
  {015503} (\bibinfo {year} {2011})}\BibitemShut {NoStop}%
\bibitem [{\citenamefont {Marqu\'es}\ \emph {et~al.}(2011)\citenamefont
  {Marqu\'es}, \citenamefont {McMahon}, \citenamefont {Gregoryanz},
  \citenamefont {Hanfland}, \citenamefont {Guillaume}, \citenamefont {Pickard},
  \citenamefont {Ackland},\ and\ \citenamefont
  {Nelmes}}]{PhysRevLett.106.095502}%
  \BibitemOpen
  \bibfield  {author} {\bibinfo {author} {\bibfnamefont {M.}~\bibnamefont
  {Marqu\'es}}, \bibinfo {author} {\bibfnamefont {M.~I.}\ \bibnamefont
  {McMahon}}, \bibinfo {author} {\bibfnamefont {E.}~\bibnamefont {Gregoryanz}},
  \bibinfo {author} {\bibfnamefont {M.}~\bibnamefont {Hanfland}}, \bibinfo
  {author} {\bibfnamefont {C.~L.}\ \bibnamefont {Guillaume}}, \bibinfo {author}
  {\bibfnamefont {C.~J.}\ \bibnamefont {Pickard}}, \bibinfo {author}
  {\bibfnamefont {G.~J.}\ \bibnamefont {Ackland}}, \ and\ \bibinfo {author}
  {\bibfnamefont {R.~J.}\ \bibnamefont {Nelmes}},\ }\href {\doibase
  10.1103/PhysRevLett.106.095502} {\bibfield  {journal} {\bibinfo  {journal}
  {Phys. Rev. Lett.}\ }\textbf {\bibinfo {volume} {106}},\ \bibinfo {pages}
  {095502} (\bibinfo {year} {2011})}\BibitemShut {NoStop}%
\bibitem [{\citenamefont {Naumov}\ \emph {et~al.}(2015)\citenamefont {Naumov},
  \citenamefont {Hemley}, \citenamefont {Hoffmann},\ and\ \citenamefont
  {Ashcroft}}]{naumov2015chemical}%
  \BibitemOpen
  \bibfield  {author} {\bibinfo {author} {\bibfnamefont {I.~I.}\ \bibnamefont
  {Naumov}}, \bibinfo {author} {\bibfnamefont {R.~J.}\ \bibnamefont {Hemley}},
  \bibinfo {author} {\bibfnamefont {R.}~\bibnamefont {Hoffmann}}, \ and\
  \bibinfo {author} {\bibfnamefont {N.~W.}\ \bibnamefont {Ashcroft}},\ }\href
  {\doibase http://dx.doi.org/10.1063/1.4928076} {\bibfield  {journal}
  {\bibinfo  {journal} {The Journal of Chemical Physics}\ }\textbf {\bibinfo
  {volume} {143}},\ \bibinfo {pages} {064702} (\bibinfo {year} {2015})},\
  \bibinfo {note} {doi: 10.1063/1.4928076}\BibitemShut {NoStop}%
\bibitem [{\citenamefont {Kokail}\ \emph {et~al.}(2016)\citenamefont {Kokail},
  \citenamefont {Heil},\ and\ \citenamefont
  {Boeri}}]{kokail_PhysRevB.94.060502}%
  \BibitemOpen
  \bibfield  {author} {\bibinfo {author} {\bibfnamefont {C.}~\bibnamefont
  {Kokail}}, \bibinfo {author} {\bibfnamefont {C.}~\bibnamefont {Heil}}, \ and\
  \bibinfo {author} {\bibfnamefont {L.}~\bibnamefont {Boeri}},\ }\href
  {\doibase 10.1103/PhysRevB.94.060502} {\bibfield  {journal} {\bibinfo
  {journal} {Phys. Rev. B}\ }\textbf {\bibinfo {volume} {94}},\ \bibinfo
  {pages} {060502} (\bibinfo {year} {2016})}\BibitemShut {NoStop}%
\bibitem [{\citenamefont {Eremets}\ \emph {et~al.}(2001)\citenamefont
  {Eremets}, \citenamefont {Struzhkin}, \citenamefont {Mao},\ and\
  \citenamefont {Hemley}}]{Eremets272}%
  \BibitemOpen
  \bibfield  {author} {\bibinfo {author} {\bibfnamefont {M.~I.}\ \bibnamefont
  {Eremets}}, \bibinfo {author} {\bibfnamefont {V.~V.}\ \bibnamefont
  {Struzhkin}}, \bibinfo {author} {\bibfnamefont {H.-k.}\ \bibnamefont {Mao}},
  \ and\ \bibinfo {author} {\bibfnamefont {R.~J.}\ \bibnamefont {Hemley}},\
  }\href {\doibase 10.1126/science.1062286} {\bibfield  {journal} {\bibinfo
  {journal} {Science}\ }\textbf {\bibinfo {volume} {293}},\ \bibinfo {pages}
  {272} (\bibinfo {year} {2001})}\BibitemShut {NoStop}%
\bibitem [{\citenamefont {Papaconstantopoulos}\ and\ \citenamefont
  {Mehl}(2002)}]{PhysRevB.65.172510}%
  \BibitemOpen
  \bibfield  {author} {\bibinfo {author} {\bibfnamefont {D.~A.}\ \bibnamefont
  {Papaconstantopoulos}}\ and\ \bibinfo {author} {\bibfnamefont {M.~J.}\
  \bibnamefont {Mehl}},\ }\href {\doibase 10.1103/PhysRevB.65.172510}
  {\bibfield  {journal} {\bibinfo  {journal} {Phys. Rev. B}\ }\textbf {\bibinfo
  {volume} {65}},\ \bibinfo {pages} {172510} (\bibinfo {year}
  {2002})}\BibitemShut {NoStop}%
\bibitem [{\citenamefont {Hu}\ \emph {et~al.}(2013)\citenamefont {Hu},
  \citenamefont {Oganov}, \citenamefont {Zhu}, \citenamefont {Qian},
  \citenamefont {Frapper}, \citenamefont {Lyakhov},\ and\ \citenamefont
  {Zhou}}]{PhysRevLett.110.165504}%
  \BibitemOpen
  \bibfield  {author} {\bibinfo {author} {\bibfnamefont {C.-H.}\ \bibnamefont
  {Hu}}, \bibinfo {author} {\bibfnamefont {A.~R.}\ \bibnamefont {Oganov}},
  \bibinfo {author} {\bibfnamefont {Q.}~\bibnamefont {Zhu}}, \bibinfo {author}
  {\bibfnamefont {G.-R.}\ \bibnamefont {Qian}}, \bibinfo {author}
  {\bibfnamefont {G.}~\bibnamefont {Frapper}}, \bibinfo {author} {\bibfnamefont
  {A.~O.}\ \bibnamefont {Lyakhov}}, \ and\ \bibinfo {author} {\bibfnamefont
  {H.-Y.}\ \bibnamefont {Zhou}},\ }\href {\doibase
  10.1103/PhysRevLett.110.165504} {\bibfield  {journal} {\bibinfo  {journal}
  {Phys. Rev. Lett.}\ }\textbf {\bibinfo {volume} {110}},\ \bibinfo {pages}
  {165504} (\bibinfo {year} {2013})}\BibitemShut {NoStop}%
\bibitem [{\citenamefont {Kolmogorov}\ and\ \citenamefont
  {Curtarolo}(2006)}]{kolmogorov_lib_PRB2006}%
  \BibitemOpen
  \bibfield  {author} {\bibinfo {author} {\bibfnamefont {A.~N.}\ \bibnamefont
  {Kolmogorov}}\ and\ \bibinfo {author} {\bibfnamefont {S.}~\bibnamefont
  {Curtarolo}},\ }\href {\doibase 10.1103/PhysRevB.73.180501} {\bibfield
  {journal} {\bibinfo  {journal} {Phys. Rev. B}\ }\textbf {\bibinfo {volume}
  {73}},\ \bibinfo {pages} {180501} (\bibinfo {year} {2006})}\BibitemShut
  {NoStop}%
\bibitem [{usp()}]{uspex_note}%
  \BibitemOpen
  \href@noop {} {}\bibinfo {note} {USPEX employs an advanced evolutionary
  algorithm technique to find the global minimum of the energy (or enthalpy)
  landscape of a given system; each point of this hyperspace is generated by an
  {\em ab-initio} calculation. An initial population of structures, generated
  randomly or according to user-specified criteria, is evolved using mechanisms
  from biological evolution (selection, mutation, reproduction, survival of the
  fitness) to minimize a {\em fitness} function, which is typically the
  internal energy (or enthalpy) of the system.}\BibitemShut {Stop}%
\bibitem [{\citenamefont {Oganov}\ and\ \citenamefont
  {Glass}(2006)}]{oganov2006crystal}%
  \BibitemOpen
  \bibfield  {author} {\bibinfo {author} {\bibfnamefont {A.~R.}\ \bibnamefont
  {Oganov}}\ and\ \bibinfo {author} {\bibfnamefont {C.~W.}\ \bibnamefont
  {Glass}},\ }\href {\doibase http://dx.doi.org/10.1063/1.2210932} {\bibfield
  {journal} {\bibinfo  {journal} {The Journal of Chemical Physics}\ }\textbf
  {\bibinfo {volume} {124}},\ \bibinfo {pages} {244704} (\bibinfo {year}
  {2006})},\ \bibinfo {note} {doi: 10.1063/1.2210932}\BibitemShut {NoStop}%
\bibitem [{\citenamefont {{Lyakhov Andriy O.}}\ \emph
  {et~al.}(2013)\citenamefont {{Lyakhov Andriy O.}}, \citenamefont {{Oganov
  Artem R.}}, \citenamefont {{Stokes Harold T.}},\ and\ \citenamefont {{Zhu
  Qiang}}}]{lyakhov2013new}%
  \BibitemOpen
  \bibfield  {author} {\bibinfo {author} {\bibnamefont {{Lyakhov Andriy O.}}},
  \bibinfo {author} {\bibnamefont {{Oganov Artem R.}}}, \bibinfo {author}
  {\bibnamefont {{Stokes Harold T.}}}, \ and\ \bibinfo {author} {\bibnamefont
  {{Zhu Qiang}}},\ }\href
  {http://www.sciencedirect.com/science/article/pii/S0010465512004055}
  {\bibfield  {journal} {\bibinfo  {journal} {Computer Physics Communications}\
  }\textbf {\bibinfo {volume} {184}},\ \bibinfo {pages} {1172} (\bibinfo {year}
  {2013})}\BibitemShut {NoStop}%
\bibitem [{\citenamefont {{Oganov Artem R.}}\ \emph {et~al.}(2011)\citenamefont
  {{Oganov Artem R.}}, \citenamefont {{Lyakhov Andriy O.}},\ and\ \citenamefont
  {{Valle Mario}}}]{oganov2011evolutionary}%
  \BibitemOpen
  \bibfield  {author} {\bibinfo {author} {\bibnamefont {{Oganov Artem R.}}},
  \bibinfo {author} {\bibnamefont {{Lyakhov Andriy O.}}}, \ and\ \bibinfo
  {author} {\bibnamefont {{Valle Mario}}},\ }\href {\doibase
  http://dx.doi.org/10.1021/ar1001318} {\bibfield  {journal} {\bibinfo
  {journal} {Accounts of Chemical Research}\ }\textbf {\bibinfo {volume}
  {44}},\ \bibinfo {pages} {227} (\bibinfo {year} {2011})},\ \bibinfo {note}
  {doi: 10.1021/ar1001318}\BibitemShut {NoStop}%
\bibitem [{CD_({\natexlab{a}})}]{CD_note1}%
  \BibitemOpen
  \href@noop {} {} ({\natexlab{a}}),\ \bibinfo {note} {for effectively sampling
  the energy landscape, we employed the antiseeds technique for all runs
  \cite{Lyakhov20131172}. The underlying structural relaxations were performed
  using the \textsc{Vasp}-code within the all electron projector augmented wave
  method~\cite{PhysRevB.47.558,PhysRevB.54.11169,PhysRevB.59.1758,PhysRevB.50.17953}.
  We used the Perdew-Burke-Ernzerhof parametrization for exchange-correlation
  potential~\cite{PhysRevLett.77.3865}. For the evolutionary runs we used a
  maximal kinetic energy cutoff of 800$\,$eV and a $\bb{k}$-point resolution of
  0.04$\,2\pi$/\AA.}\BibitemShut {Stop}%
\bibitem [{CD_({\natexlab{b}})}]{CD_NOTE2}%
  \BibitemOpen
  \href@noop {} {} ({\natexlab{b}}),\ \bibinfo {note} {for the most stable
  structures along the binary lines, we carried out restricted structural
  relaxations with plane wave cutoffs between 1000 and 1200$\,$eV, with a
  $\bb{k}$-point resolution of 0.01$\,2\pi$/\AA, which ensured converge of the
  total energy better than 1$\,$meV per atom. Moreover, we ensured that all
  stress-tensors are diagonal and isotropic with fluctuations in the diagonal
  elements smaller than 0.1$\,$kbar. All ionic forces are converged to values
  better than 1$\,$meV/\AA.}\BibitemShut {Stop}%
\bibitem [{\citenamefont {Matthias}\ \emph {et~al.}(1968)\citenamefont
  {Matthias}, \citenamefont {Geballe}, \citenamefont {Andres}, \citenamefont
  {Corenzwit}, \citenamefont {Hull},\ and\ \citenamefont
  {Maita}}]{Matthias_B12}%
  \BibitemOpen
  \bibfield  {author} {\bibinfo {author} {\bibfnamefont {B.~T.}\ \bibnamefont
  {Matthias}}, \bibinfo {author} {\bibfnamefont {T.~H.}\ \bibnamefont
  {Geballe}}, \bibinfo {author} {\bibfnamefont {K.}~\bibnamefont {Andres}},
  \bibinfo {author} {\bibfnamefont {E.}~\bibnamefont {Corenzwit}}, \bibinfo
  {author} {\bibfnamefont {G.~W.}\ \bibnamefont {Hull}}, \ and\ \bibinfo
  {author} {\bibfnamefont {J.~P.}\ \bibnamefont {Maita}},\ }\href {\doibase
  10.1126/science.159.3814.530} {\bibfield  {journal} {\bibinfo  {journal}
  {Science}\ }\textbf {\bibinfo {volume} {159}},\ \bibinfo {pages} {530}
  (\bibinfo {year} {1968})}\BibitemShut {NoStop}%
\bibitem [{\citenamefont {Peles}\ \emph {et~al.}(2004)\citenamefont {Peles},
  \citenamefont {Alford}, \citenamefont {Ma}, \citenamefont {Yang},\ and\
  \citenamefont {Chou}}]{Peles_PRB_2004}%
  \BibitemOpen
  \bibfield  {author} {\bibinfo {author} {\bibfnamefont {A.}~\bibnamefont
  {Peles}}, \bibinfo {author} {\bibfnamefont {J.~A.}\ \bibnamefont {Alford}},
  \bibinfo {author} {\bibfnamefont {Z.}~\bibnamefont {Ma}}, \bibinfo {author}
  {\bibfnamefont {L.}~\bibnamefont {Yang}}, \ and\ \bibinfo {author}
  {\bibfnamefont {M.~Y.}\ \bibnamefont {Chou}},\ }\href {\doibase
  10.1103/PhysRevB.70.165105} {\bibfield  {journal} {\bibinfo  {journal} {Phys.
  Rev. B}\ }\textbf {\bibinfo {volume} {70}},\ \bibinfo {pages} {165105}
  (\bibinfo {year} {2004})}\BibitemShut {NoStop}%
\bibitem [{\citenamefont {Huan}\ \emph {et~al.}(2013)\citenamefont {Huan},
  \citenamefont {Amsler}, \citenamefont {Marques}, \citenamefont {Botti},
  \citenamefont {Willand},\ and\ \citenamefont {Goedecker}}]{marques_PRL_2013}%
  \BibitemOpen
  \bibfield  {author} {\bibinfo {author} {\bibfnamefont {T.~D.}\ \bibnamefont
  {Huan}}, \bibinfo {author} {\bibfnamefont {M.}~\bibnamefont {Amsler}},
  \bibinfo {author} {\bibfnamefont {M.~A.~L.}\ \bibnamefont {Marques}},
  \bibinfo {author} {\bibfnamefont {S.}~\bibnamefont {Botti}}, \bibinfo
  {author} {\bibfnamefont {A.}~\bibnamefont {Willand}}, \ and\ \bibinfo
  {author} {\bibfnamefont {S.}~\bibnamefont {Goedecker}},\ }\href {\doibase
  10.1103/PhysRevLett.110.135502} {\bibfield  {journal} {\bibinfo  {journal}
  {Phys. Rev. Lett.}\ }\textbf {\bibinfo {volume} {110}},\ \bibinfo {pages}
  {135502} (\bibinfo {year} {2013})}\BibitemShut {NoStop}%
\bibitem [{\citenamefont {McMillan}(1968)}]{McMillanTC}%
  \BibitemOpen
  \bibfield  {author} {\bibinfo {author} {\bibfnamefont {W.~L.}\ \bibnamefont
  {McMillan}},\ }\href {\doibase 10.1103/PhysRev.167.331} {\bibfield  {journal}
  {\bibinfo  {journal} {Phys. Rev.}\ }\textbf {\bibinfo {volume} {167}},\
  \bibinfo {pages} {331} (\bibinfo {year} {1968})}\BibitemShut {NoStop}%
\bibitem [{\citenamefont {Allen}\ and\ \citenamefont
  {Dynes}(1975)}]{AllenDynes_PRB1975}%
  \BibitemOpen
  \bibfield  {author} {\bibinfo {author} {\bibfnamefont {P.~B.}\ \bibnamefont
  {Allen}}\ and\ \bibinfo {author} {\bibfnamefont {R.~C.}\ \bibnamefont
  {Dynes}},\ }\href {\doibase 10.1103/PhysRevB.12.905} {\bibfield  {journal}
  {\bibinfo  {journal} {Phys. Rev. B}\ }\textbf {\bibinfo {volume} {12}},\
  \bibinfo {pages} {905} (\bibinfo {year} {1975})}\BibitemShut {NoStop}%
\bibitem [{\citenamefont {Ge}\ \emph {et~al.}(2016)\citenamefont {Ge},
  \citenamefont {Zhang},\ and\ \citenamefont {Yao}}]{Ge_PRB_2016}%
  \BibitemOpen
  \bibfield  {author} {\bibinfo {author} {\bibfnamefont {Y.}~\bibnamefont
  {Ge}}, \bibinfo {author} {\bibfnamefont {F.}~\bibnamefont {Zhang}}, \ and\
  \bibinfo {author} {\bibfnamefont {Y.}~\bibnamefont {Yao}},\ }\href {\doibase
  10.1103/PhysRevB.93.224513} {\bibfield  {journal} {\bibinfo  {journal} {Phys.
  Rev. B}\ }\textbf {\bibinfo {volume} {93}},\ \bibinfo {pages} {224513}
  (\bibinfo {year} {2016})}\BibitemShut {NoStop}%
\bibitem [{\citenamefont {Lyakhov}\ \emph {et~al.}(2013)\citenamefont
  {Lyakhov}, \citenamefont {Oganov}, \citenamefont {Stokes},\ and\
  \citenamefont {Zhu}}]{Lyakhov20131172}%
  \BibitemOpen
  \bibfield  {author} {\bibinfo {author} {\bibfnamefont {A.~O.}\ \bibnamefont
  {Lyakhov}}, \bibinfo {author} {\bibfnamefont {A.~R.}\ \bibnamefont {Oganov}},
  \bibinfo {author} {\bibfnamefont {H.~T.}\ \bibnamefont {Stokes}}, \ and\
  \bibinfo {author} {\bibfnamefont {Q.}~\bibnamefont {Zhu}},\ }\href {\doibase
  http://dx.doi.org/10.1016/j.cpc.2012.12.009} {\bibfield  {journal} {\bibinfo
  {journal} {Computer Physics Communications}\ }\textbf {\bibinfo {volume}
  {184}},\ \bibinfo {pages} {1172 } (\bibinfo {year} {2013})}\BibitemShut
  {NoStop}%
\bibitem [{\citenamefont {Kresse}\ and\ \citenamefont
  {Hafner}(1993)}]{PhysRevB.47.558}%
  \BibitemOpen
  \bibfield  {author} {\bibinfo {author} {\bibfnamefont {G.}~\bibnamefont
  {Kresse}}\ and\ \bibinfo {author} {\bibfnamefont {J.}~\bibnamefont
  {Hafner}},\ }\href {\doibase 10.1103/PhysRevB.47.558} {\bibfield  {journal}
  {\bibinfo  {journal} {Phys. Rev. B}\ }\textbf {\bibinfo {volume} {47}},\
  \bibinfo {pages} {558} (\bibinfo {year} {1993})}\BibitemShut {NoStop}%
\bibitem [{\citenamefont {Kresse}\ and\ \citenamefont
  {Furthm\"uller}(1996)}]{PhysRevB.54.11169}%
  \BibitemOpen
  \bibfield  {author} {\bibinfo {author} {\bibfnamefont {G.}~\bibnamefont
  {Kresse}}\ and\ \bibinfo {author} {\bibfnamefont {J.}~\bibnamefont
  {Furthm\"uller}},\ }\href {\doibase 10.1103/PhysRevB.54.11169} {\bibfield
  {journal} {\bibinfo  {journal} {Phys. Rev. B}\ }\textbf {\bibinfo {volume}
  {54}},\ \bibinfo {pages} {11169} (\bibinfo {year} {1996})}\BibitemShut
  {NoStop}%
\bibitem [{\citenamefont {Kresse}\ and\ \citenamefont
  {Joubert}(1999)}]{PhysRevB.59.1758}%
  \BibitemOpen
  \bibfield  {author} {\bibinfo {author} {\bibfnamefont {G.}~\bibnamefont
  {Kresse}}\ and\ \bibinfo {author} {\bibfnamefont {D.}~\bibnamefont
  {Joubert}},\ }\href {\doibase 10.1103/PhysRevB.59.1758} {\bibfield  {journal}
  {\bibinfo  {journal} {Phys. Rev. B}\ }\textbf {\bibinfo {volume} {59}},\
  \bibinfo {pages} {1758} (\bibinfo {year} {1999})}\BibitemShut {NoStop}%
\bibitem [{\citenamefont {Bl\"ochl}(1994)}]{PhysRevB.50.17953}%
  \BibitemOpen
  \bibfield  {author} {\bibinfo {author} {\bibfnamefont {P.~E.}\ \bibnamefont
  {Bl\"ochl}},\ }\href {\doibase 10.1103/PhysRevB.50.17953} {\bibfield
  {journal} {\bibinfo  {journal} {Phys. Rev. B}\ }\textbf {\bibinfo {volume}
  {50}},\ \bibinfo {pages} {17953} (\bibinfo {year} {1994})}\BibitemShut
  {NoStop}%
\bibitem [{\citenamefont {Perdew}\ \emph {et~al.}(1996)\citenamefont {Perdew},
  \citenamefont {Burke},\ and\ \citenamefont
  {Ernzerhof}}]{PhysRevLett.77.3865}%
  \BibitemOpen
  \bibfield  {author} {\bibinfo {author} {\bibfnamefont {J.~P.}\ \bibnamefont
  {Perdew}}, \bibinfo {author} {\bibfnamefont {K.}~\bibnamefont {Burke}}, \
  and\ \bibinfo {author} {\bibfnamefont {M.}~\bibnamefont {Ernzerhof}},\ }\href
  {\doibase 10.1103/PhysRevLett.77.3865} {\bibfield  {journal} {\bibinfo
  {journal} {Phys. Rev. Lett.}\ }\textbf {\bibinfo {volume} {77}},\ \bibinfo
  {pages} {3865} (\bibinfo {year} {1996})}\BibitemShut {NoStop}%
\end{thebibliography}
\end{document}